\documentclass[nofootinbib,prd,superscriptaddress,twocolumn]{revtex4}
\usepackage{amsmath,graphicx,hyperref}
\usepackage{slashed}
\usepackage{feynmp}
\usepackage{epstopdf}
\usepackage{color}
\usepackage[dvipsnames]{xcolor}

\newcommand{\be}{\begin{equation}} 
\newcommand{\ee}{\end{equation}} 
\newcommand{\beq}{\begin{equation}} 
\newcommand{\eeq}{\end{equation}} 
\newcommand{\bea}{\begin{eqnarray}} 
\newcommand{\eea}{\end{eqnarray}}

\newcommand{\qa}{q_{\mathrm{A}'}}

\newcommand{\Aprime}{\mathrm{A}'}
\newcommand{\rdp}{r_{\mathrm{dp}}}

\newcommand{\dd}{\mathrm{d}}

\begin{document}

\baselineskip 3.0ex 

\vspace*{18pt}


\title{Updated Constraints on Self-Interacting Dark Matter from Supernova 1987A}

\def\Pitt{Pittsburgh Particle Physics Astrophysics and Cosmology Center (PITT PACC) \\ Department of Physics and Astronomy, University of Pittsburgh, Pittsburgh, Pennsylvania 15260, USA}

\author{Cameron Mahoney}
\email[E-mail:]{cbm34@pitt.edu}
\affiliation{\Pitt}
\author{Adam K. Leibovich}
\email[E-mail:]{akl2@pitt.edu}
\affiliation{\Pitt}
\author{Andrew R. Zentner}
\email[E-mail:]{zentner@pitt.edu}
\affiliation{\Pitt}  

\begin{abstract} 
\baselineskip 3.0ex   

We revisit SN1987A constraints on light, hidden sector gauge bosons (``dark photons'') that are coupled to the standard model through kinetic mixing with the photon. These constraints are realized because excessive bremsstrahlung radiation of the dark photon can lead to rapid cooling of the SN1987A progenitor core, in contradiction to the observed neutrinos from that event. The models we consider are of interest as phenomenological models of strongly self-interacting dark matter. We clarify several possible ambiguities in the literature and identify errors in prior analyses. We find constraints on the dark photon mixing parameter that are in rough agreement with the early estimates of Dent et al. \cite{dent_etal12}, but only because significant errors in their analyses fortuitously canceled. Our constraints are in good agreement with subsequent analyses by Rrapaj \& Reddy \cite{rrapaj_reddy16} and Hardy \& Lasenby \cite{hardy_lasenby17}. We estimate the dark photon bremsstrahlung rate using one-pion exchange (OPE), while Rrapaj \& Reddy use a soft radiation approximation (SRA) to exploit measured nuclear scattering cross sections. We find that the differences between mixing parameter constraints obtained through the OPE approximation or the SRA approximation are roughly a factor of $\sim 2-3$. Hardy \& Laseby \cite{hardy_lasenby17} include plasma effects in their calculations finding significantly weaker constraints on dark photon mixing for dark photon masses below $\sim 10\, \mathrm{MeV}$. We do not consider plasma effects. Lastly, we point out that the properties of the SN1987A progenitor core remain somewhat uncertain and that this uncertainty alone causes uncertainty of at least a factor of $\sim 2-3$ in the excluded values of the dark photon mixing parameter. Further refinement of these estimates is unwarranted until either the interior of the SN1987A progenitor is more well understood or additional, large, and heretofore neglected effects, such as the plasma interactions studied by Hardy \& Lasenby \cite{hardy_lasenby17}, are identified.
\end{abstract}

\maketitle

\section{Introduction}

An overwhelming preponderance of observational evidence indicates that a form of nonrelativistic, nonbaryonic, 
dark matter constitutes the majority of mass in the Universe and drives the formation of cosmic structure. 
The pace of the quest to identify the dark matter is accelerating on many fronts. Weakly-interacting massive 
particles (WIMPs) have received the most attention as dark matter candidates (see Ref.~\cite{jungman_etal96} for a review). 
Dark matter particles that interact with standard model particles only weakly, while interacting among themselves 
much more strongly have been studied as an alternative to WIMP scenarios in numerous contexts 
\cite{carlson_etal92,deLaix_etal95,atrio-barandela_davidson97,spergel_steinhardt00,hogan_dalcanton00,mohapatra_teplitz00,
dave_etal01,hisano_etal04,hisano_etal05,pospelov_etal08,arkani-hamed_etal08a,lattanzi_silk08,ackerman_etal09,feng_etal09,
kong_etal15,Buckley:2009in,Boddy:2014yra,Boddy:2014qxa,Boddy:2016bbu}
and constraints on self-interacting dark matter (SIDM) models have been explored by many authors \cite{yoshida_etal00,gnedin_ostriker01,miralda-escude02,randall_etal08,kamionkowski_profumo08,zentner09,robertson_zentner09,pieri_etal09,spolyar_etal09,finkbeiner_etal09,slatyer_etal09,bramante_etal14,albuquerque_etal14,kaplinghat_etal14,chen_etal14,feng_etal16,catena_widmark16,Markevitch:2003at,Zavala:2012us,Rocha:2012jg,Peter:2012jh,Kahlhoefer:2013dca,Elbert:2014bma,Feng:2015hja,DelNobile:2015uua,Feng:2016ijc,Dooley:2016ajo,Kim:2016ujt,Bringmann:2016din}.
In this paper, we revisit and update astrophysical constraints on SIDM models from supernova cooling.

SIDM models in which large self-interaction cross sections are mediated by sufficiently light 
bosons ($M \lesssim 100$~GeV) can be constrained astrophysically using supernovae, particularly 
SN1987A. Light gauge bosons will be produced within the hot supernova core, primarily through 
bremsstrahlung, and radiated. This non-standard energy loss mechanism can result in an energy loss rate 
from the supernova core that is inconsistent with observations of neutrinos from SN1987A, 
analogous to the classic constraint on axions \cite{turner88,raffelt96_book}. 
This effect has already been exploited by Dent et al. \cite{dent_etal12}, Rrapaj and Reddy \cite{rrapaj_reddy16}, and 
Hardy and Lasenby \cite{hardy_lasenby17} 
to constraint dark electromagnetism models in which the new gauge boson, the so-called dark photon, 
is kinetically mixed with the standard model photon. See also \cite{Kazanas:2014mca,Zhang:2014wra,Chang:2016ntp}. The SN1987A constraint places a limit on the mixing parameter.

We initiated our study because of a number of ambiguities appearing in the previous literature on this subject. 
In particular, we could not reproduce the constraints of Ref.~\cite{dent_etal12}. During the course of our 
study, we identified a number of errors in the analysis of Ref.~\cite{dent_etal12}. First, it is straightforward 
to demonstrate that the kinematical relationships given in Appendix A of Ref.~\cite{dent_etal12} are 
incorrect. Second, the squared matrix elements given in Eq.~(A3) and Eq.~(C18) of Ref.~\cite{dent_etal12} must 
be incorrect. These matrix elements do not obey the correct symmetries under interchange of incoming 
and/or outgoing momenta. Furthermore, Ref.~\cite{dent_etal12} neglects the mass of the gauge boson, which 
is legitimate in the classic case of the $\sim$meV-mass axion, but not in the present context. Finally, Ref.~\cite{dent_etal12} 
employed an inconsistent model for the permitted energy loss rate from the supernova interior. Our work amounts 
primarily to repeating the calculation of Ref.~\cite{dent_etal12} in order to rectify these oversights. Our primary 
calculation treats nucleon scattering via one-pion exchange (OPE).

As we were completing our manuscript, two related papers were published. Rrapaj and Reddy \cite{rrapaj_reddy16} 
computed bounds on mixing of the dark and standard model photons using a soft radiation approximation (SRA) 
for dark photon bremsstrahlung. This has the distinct advantage of enabling nucleon scattering data to be used 
directly in the estimation of the bremsstrahlung rate, but is only approximate because at large dark photon masses 
the radiated dark photons carry off considerable momentum and energy. We have been able to reproduce the 
result of Rrapaj and Reddy \cite{rrapaj_reddy16} and find that the SRA plausibly results in only a factor of 
$\sim 3$ underestimate of the upper bound on the dark photon mixing parameter at most. Moreover, 
our OPE results agree well with the SRA calculation of Rrapaj and Reddy \cite{rrapaj_reddy16}. Nonetheless, 
Ref.~\cite{rrapaj_reddy16} quote results very similar to those of Dent et al. \cite{dent_etal12}. We find, 
rather remarkably, that the various errors in the analysis of Ref.~\cite{dent_etal12} conspire to yield a constraint that 
is very similar to the correct answer.

More recently, Hardy and Lasenby \cite{hardy_lasenby17} studied bounds on these same models (and others) 
including plasma effects. For simplicity, we have not included these plasma effects in our calculations. Hardy 
and Lasenby base their calculation off of the SRA of Ref.~\cite{rrapaj_reddy16}. Consequently, 
for dark photon masses $\gtrsim 10$~MeV they find very similar results to Ref.~\cite{rrapaj_reddy16} as well 
as the constraints we present in this manuscript. For dark photon masses $\lesssim 10$~MeV, Hardy and 
Lasenby demonstrate that constraints on the dark photon mixing parameter are significantly weaker than 
one finds when neglecting plasma effects \cite{hardy_lasenby17}.

It is important to delineate correctly the range of the viable parameter space for SIDM models. 
The parameter space available to dark electromagnetism models of SIDM has been studied extensively 
not only in the aforementioned papers, but also in the work of Bjorken et al. \cite{bjorken_etal09} and 
the Snowmass white paper by Kaplinghat, Tulin, and Yu \cite{kaplinghat_etal13_whitepaper}. One 
point that is clear from previous work is that there is at most only a slim sliver of parameter space that 
can simultaneously yield the correct relic abundance of SIDM through thermal production, 
have interesting effects on cosmological structure formation, and 
evade all constraints including the constraints from SN1987A. 
In accord with Ref.~\cite{rrapaj_reddy16} and Ref.~\cite{hardy_lasenby17}, we find that 
the constraints quoted in Ref.~\cite{dent_etal12} are too restrictive, but only by a factor 
of $\sim 4$ due to a conspiratorial cancellation of errors in Ref.~\cite{dent_etal12}. 
Hardy and Lasenby go on to demonstrate that these constraints are significantly too 
restrictive for dark photon masses $\lesssim 10$~MeV due to plasma effects.

The remainder of this paper is organized as follows. In Section~\ref{section:model}, we discuss 
dark photon models. We describe our calculation of SN1987A constraints on SIDM in 
Section~\ref{section:computational} and present our primary results in Section~\ref{section:results}. 
We stress only those points that are key to understanding the relationship between our work 
and the work of both Dent et al. in Ref.~\cite{dent_etal12} and Rrapaj and Reddy in Ref.~\cite{rrapaj_reddy16}. 
We summarize our work and draw conclusions in Section~\ref{section:conclusions}.

\section{Dark Photon Model of SIDM}
\label{section:model}

We consider constraints on SIDM specifically within the context of dark electromagnetism models. 
Dark electromagnetism models are models in which a hidden, dark sector contains a broken U(1)$'$ 
symmetry and the U(1)$'$ gauge boson is kinetically mixed with the standard model photon. For the 
purposes of this study, this is important because it demands that the Lagrangian contains terms such as 
\begin{equation}
\mathcal{L}_{\mathrm{int}} = g_{\chi} \bar{\chi} \, \widetilde{\slashed{\Aprime}} \chi + q \bar{f}\, \widetilde{\slashed{A}} f, 
\end{equation}
where $\chi$ is the dark matter, $g_{\chi}$ is the dark coupling, $\widetilde{\Aprime}$ is the dark gauge boson, 
$f$ is a standard model fermion of charge $q$, and $\widetilde{A}$ is the standard model gauge boson. 
The kinetic mixing, through a term 
$\frac{1}{2}\frac{\varepsilon}{\sqrt{1+\varepsilon^2}}\, \widetilde{F}_{\mu \nu}\widetilde{F}'^{\mu \nu}$ 
in the Lagrangian causes the $\widetilde{A}$ to be an admixture of the massless photon $A$, 
and the dark photon $\Aprime$, 
of mass $m_{\mathrm{\Aprime}} = m_{\mathrm{\widetilde{A}'}} \sqrt{1 + \varepsilon^2} \simeq m_\mathrm{{\widetilde{A}'}}$ because 
the viable parameter range has $\varepsilon \ll 1$. The dark matter particles are thereby coupled 
to the standard model fermions with a coupling constant $\varepsilon q$, 
where $\varepsilon$ is the kinetic mixing parameter. The first term in this 
interaction Lagrangian gives rise to the dark matter self-interactions. 

Dark gauge bosons are produced in astrophysical environments such as supernova 
cores primarily via bremsstrahlung off of standard model particles. This bremsstrahlung 
occurs through the $\varepsilon q$ coupling to charged standard model particles, 
in this  particular case the proton and pion. The rate of bremsstrahlung 
depends upon both $\varepsilon$ and the mass of the $\Aprime$. Consequently, 
supernova cooling can constrain the mixing $\varepsilon$ as a function of 
$m_{\Aprime}$ for such models. Delineating such a constraint is the 
primary aim of this paper.

\section{Methods}
\label{section:computational}
	
We aim to estimate the rate of energy loss from the core of a supernova from $\Aprime$ bremsstrahlung 
during nucleon-nucleon interactions. The calculation is analogous to the well-known estimate of axion 
emission from supernova cores described in Ref.~\cite{raffelt96_book} and references therein, but is 
more complicated because the mass of the $\Aprime$, unlike the mass of the axion, is not necessarily 
negligible. This section describes the calculation of the rate of energy loss from a supernova core 
from $\Aprime$ bremsstrahlung. 

The bremsstrahlung process is not the only process with which we must be concerned. Clearly, the rate of 
bremsstrahlung will increase with $\varepsilon$; however, $\varepsilon$ can become sufficiently large 
that the radiated gauge bosons do not escape the supernova. This happens if the $\Aprime$ particles 
either decay to or interact with standard model particles prior to exiting the supernova core. 
In either case, the energy is not lost and the $\Aprime$ does not provide a cooling channel for the supernova. 
Consequently, for a given $m_{\Aprime}$, there is a maximum $\varepsilon$ that can be constrained in this 
manner. We estimate $\Aprime$ decay and scattering probabilities, and the upper limits on the $\varepsilon$ 
constraints in this section as well. However, we note that terrestrial experiments generally rule out 
mixing parameters higher than the upper limits of the SN1987A forbidden region, so a precise estimate 
of these upper limits is not necessary.

\subsection{Bremsstrahlung of $\Aprime$ Bosons}

\begin{figure}
\includegraphics[width=8cm]{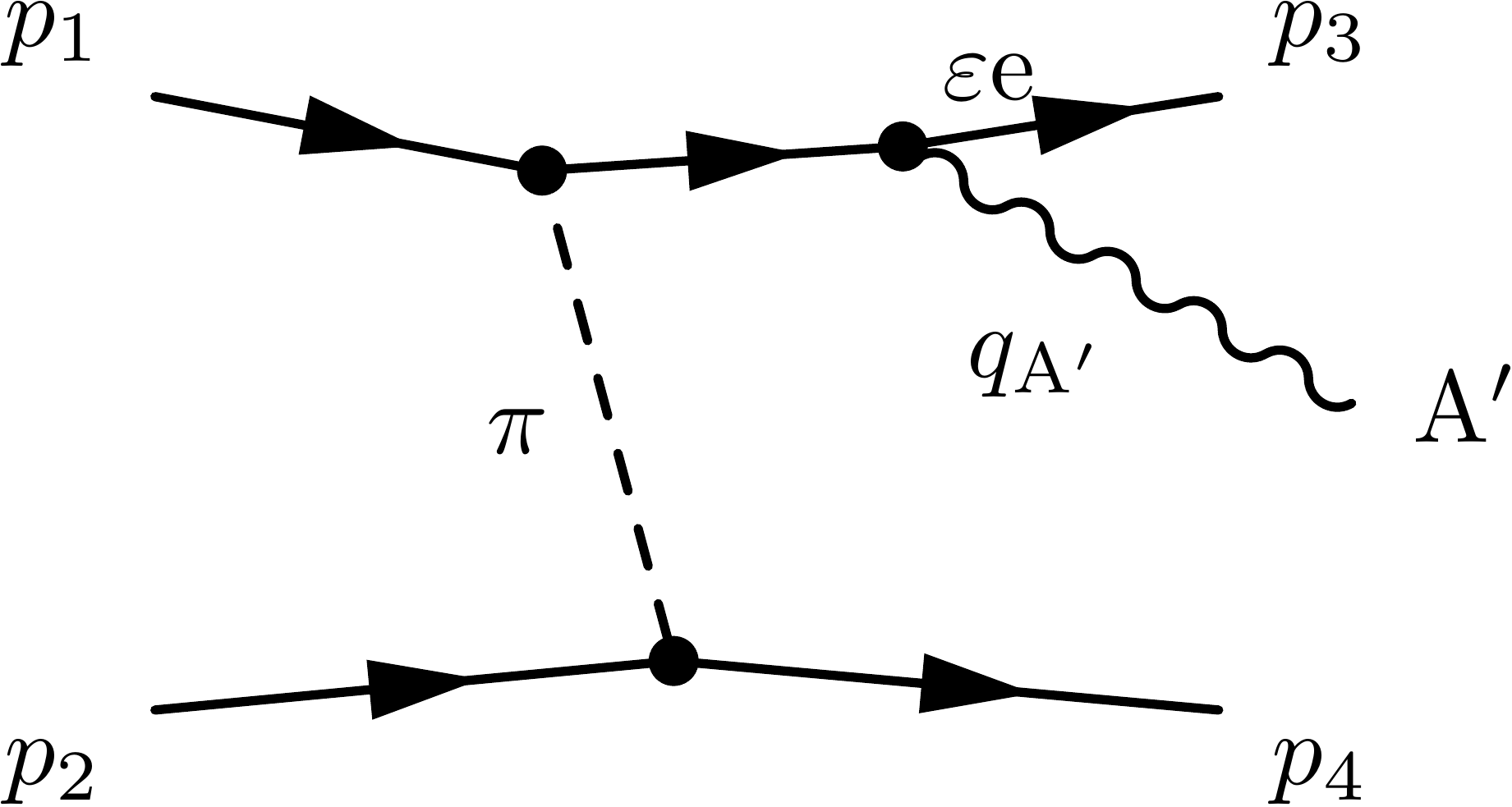}
\caption{One of the eight Feynman diagrams for the pp process. Three of the other diagrams are obtained by 
placing the $\Aprime$ on each of the protons in turn. The remaining four diagrams come from swapping the outgoing 
momenta.}
\label{fig:ppdiagram}
\end{figure}

\begin{figure}[b]
\includegraphics[width=8cm]{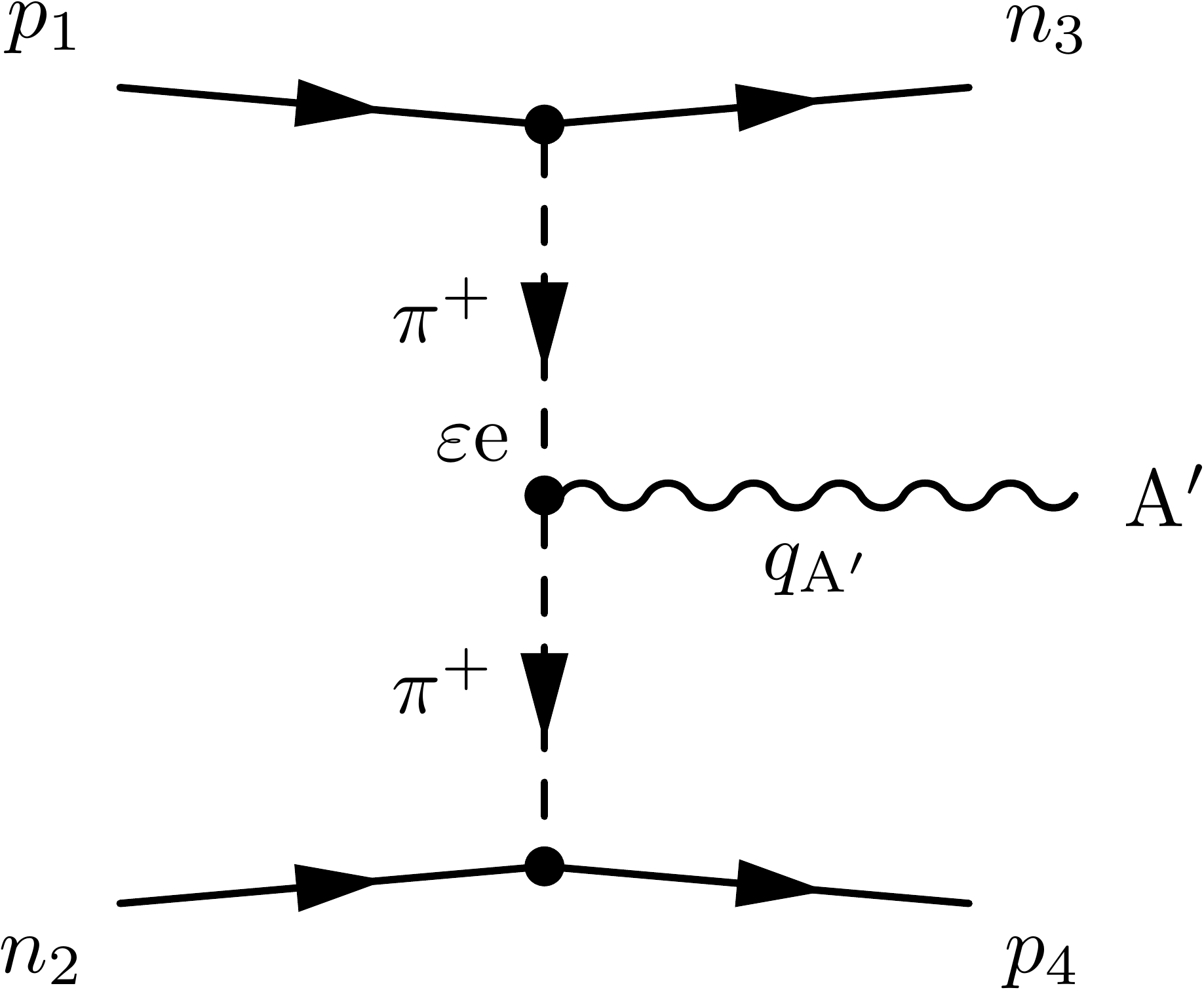}
\caption{One of the five Feynman diagrams for the pn process. This particular diagram shows 
internal bremsstrahlung off of the charged pion. The remaining four diagrams are analogous to the 
pp diagram shown in Fig.~\ref{fig:ppdiagram}. In the case of the pn processes, there are only four 
diagrams for bremsstrahlung off of the external legs because two of the legs correspond to the 
uncharged neutron.}
\label{fig:npdiagram}
\end{figure}

There are two processes to consider in order to estimate the rate of energy loss via $\Aprime$ bremsstrahlung. The first is  
proton-proton (pp) scattering with the bremsstrahlung of the dark photon off the proton; $p+p \rightarrow p+p+\Aprime$. 
The second is proton-neutron (pn) scattering with bremsstrahlung off of either the proton or the charged pion; 
$p+n \rightarrow p+n+\Aprime$. We estimate the rates for these processes using the one-pion exchange (OPE) 
approximation for nucleon interactions. In the pp case, there are eight tree-level diagrams, 
with the emission of the $\Aprime$ from each of the external legs. One of these diagrams is shown 
in Fig.~\ref{fig:ppdiagram}; the remaining seven diagrams come from placing the radiated $\Aprime$ 
on each of the other three protons and then, for each of these, interchanging the outgoing momenta. 
For the pn case, there are five diagrams, four of which are analogous to the pp diagram shown in 
Fig.~\ref{fig:ppdiagram}. The fifth diagram, shown in Fig.~\ref{fig:npdiagram}, corresponds to 
emission of the $\Aprime$ from the exchanged, charged pion.

Evaluating these diagrams is tedious, but very straightforward. The calculation differs 
from the well-known axion bremsstrahlung calculation exploited in a similar context \cite{raffelt96_book}, 
because the mass of the $\Aprime$ boson is not necessarily negligible in the kinematic 
region of interest for supernova explosions. The correct kinematical relations are 
\bea 
p_1 \cdot p_2 &=& M_{\mathrm N}^2 - \frac{l^2}{2} - \frac{k^2}{2} + p_2 \cdot \qa,\\
p_1 \cdot p_3 &=& M_{\mathrm N}^2 + k \cdot l - \frac{k^2}{2} + p_3 \cdot \qa,\\  
p_1 \cdot p_4 &=& k \cdot l + M_{\mathrm N}^2 - \frac{l^2}{2} + p_4 \cdot \qa, \\
p_2 \cdot p_3 &=& M_{\mathrm N}^2 - \frac{l^2}{2}, \\ 
p_2 \cdot p_4 &=& M_{\mathrm N}^2 - \frac{k^2}{2},\quad \mathrm{and}\\
p_3 \cdot p_4 &=& k \cdot l + M_{\mathrm N}^2 - \frac{l^2 + k^2}{2},
\eea
where $p_1$ and $p_2$ are the four-momenta of the incoming nucleons, $p_3$ and $p_4$ are the momenta 
of the outgoing nucleons, $\qa$ is the $\Aprime$ momentum, $k=p_2 - p_4$, $l=p_2 - p_3$, and $M_{\mathrm{N}}$ 
is the nucleon mass. These kinematical relations correct the relations in Ref.~\cite{dent_etal12}.

The eight diagrams contribute the following eight terms to the pp amplitude, 
\begin{widetext}
\begin{eqnarray}
	M_1 &=& \frac{4 M_{\mathrm N}^2}{ m_\pi^2} \frac{f_{\mathrm{pp}}^2\, e\, \varepsilon}{k^2-m_\pi^2}  \frac{1}{m_{\mathrm{\Aprime}}^2 - 2 \qa \cdot p_1}\ 
	\bar{u}(p_4) \gamma_5 u(p_2)\  \bar{u}(p_3) \gamma_5 (\slashed{p}_1 - \qa +M_{\mathrm N})\slashed \epsilon  u(p_1), \\
	M_2 &=& -\frac{4 M_{\mathrm N}^2}{ m_\pi^2} \frac{f_{\mathrm{pp}}^2\, e\, \varepsilon}{l^2-m_\pi^2}  \frac{1}{m_{\mathrm{\Aprime}}^2 - 2 \qa \cdot p_1}\  
	\bar{u}(p_3) \gamma_5 u(p_2)\  \bar{u}(p_4) \gamma_5 (\slashed{p}_1 - \qa +M_{\mathrm N})\slashed{\epsilon} u(p_1),\\
	M_3 &=& \frac{4 M_{\mathrm N}^2}{ m_\pi^2} \frac{f_{\mathrm{pp}}^2\, e\, \varepsilon}{k^2-m_\pi^2}  \frac{1}{m_{\mathrm{\Aprime}}^2 - 2 \qa \cdot p_2}\ 
	\bar{u}(p_3) \gamma_5 u(p_1)\ \bar{u}(p_4) \gamma_5 (\slashed{p}_2 - \qa +M_{\mathrm N})\slashed \epsilon  u(p_2), \\
	M_4 &=& -\frac{4 M_{\mathrm N}^2}{ m_\pi^2} \frac{f_{\mathrm{pp}}^2\, e\, \varepsilon}{l^2-m_\pi^2}  \frac{1}{m_{\mathrm{\Aprime}}^2 - 2 \qa \cdot p_2}\ 
	\bar{u}(p_4) \gamma_5 u(p_1)\ \bar{u}(p_3) \gamma_5 (\slashed{p}_2 - \qa +M_{\mathrm N})\slashed \epsilon  u(p_2), \\
	M_5 &=& \frac{4 M_{\mathrm N}^2}{ m_\pi^2} \frac{f_{\mathrm{pp}}^2\, e\, \varepsilon}{k^2-m_\pi^2}  \frac{1}{m_{\mathrm{\Aprime}}^2 + 2 \qa \cdot p_3}\
	\bar{u}(p_4) \gamma_5 u(p_2)\ \bar{u}(p_3) \slashed{\epsilon} (\slashed{p}_3 + \qa +M_{\mathrm N})\gamma_5 u(p_1),\\
	M_6 &=& -\frac{4 M_{\mathrm N}^2}{ m_\pi^2} \frac{f_{\mathrm{pp}}^2\, e\, \varepsilon}{l^2-m_\pi^2}  \frac{1}{m_{\mathrm{\Aprime}}^2 + 2 \qa \cdot p_3}\ 
	\bar{u}(p_4) \gamma_5 u(p_1)\ \bar{u}(p_3) \slashed{\epsilon} (\slashed{p}_3 + \qa +M_{\mathrm N})\gamma_5 u(p_2),\\
	M_7 &=& \frac{4 M_{\mathrm N}^2}{ m_\pi^2} \frac{f_{\mathrm{pp}}^2\, e\, \varepsilon}{k^2-m_\pi^2}  \frac{1}{m_{\mathrm{\Aprime}}^2 + 2 \qa \cdot p_4}\ 
	\bar{u}(p_3) \gamma_5 u(p_1)\ \bar{u}(p_4) \slashed{\epsilon} (\slashed{p}_4 + \qa +M_{\mathrm N})\gamma_5 u(p_2),\\
	M_8 &=& -\frac{4 M_{\mathrm N}^2}{ m_\pi^2} \frac{f_{\mathrm{pp}}^2\, e\, \varepsilon}{l^2-m_\pi^2}  \frac{1}{m_{\mathrm{\Aprime}}^2 + 2 \qa \cdot p_4}\ 
	\bar{u}(p_3) \gamma_5 u(p_2)\ \bar{u}(p_4) \slashed{\epsilon} (\slashed{p}_4 + \qa +M_{\mathrm N})\gamma_5 u(p_1),
\end{eqnarray}
\end{widetext}
where the dark photon polarization is given by $\epsilon^{\nu}$. These expressions are identical to those 
given in Ref.~\cite{dent_etal12}; however, they do not simplify significantly if the correct kinematics are used. 
Our squared matrix element contains over 200 terms, so we do not reproduce it here for reasons of convenience. However, we note that our result for $\vert \mathcal{M}_{\mathrm{pp}} \vert^2$ is symmetric under exchange of $k$ and $l$ as required. 

The pn process contains four diagrams analogous to the diagrams for the pp process (only four, of course, 
because the neutrons cannot radiate the $\Aprime$). The new diagram that is relevant in the pn process is shown in 
Fig.~\ref{fig:npdiagram} and yields a contribution of 
\begin{widetext}
\begin{equation}
M'_5 = \frac{4 M_{\mathrm N}}{ m_\pi} \frac{f_{pn}^2 e\, \varepsilon}{l^2-m_\pi^2}  \frac{1}{(l-\qa)^2 - m_\pi^2} 
\bar{u}(p_3) \gamma_5 u(p_1) \bar{u}(p_4) \gamma_5 u(p_2) (\qa - 2l)\cdot \epsilon.
\end{equation}
\end{widetext}
The pn processes likewise yields a squared amplitude, 
$\vert \mathcal{M}_{\mathrm{pn}} \vert^2$ that is unwieldy, so we do not give the 
the full expression here. We provide mathematica notebooks detailing our 
computation of the amplitudes with this submission.

\subsection{The Streaming Limit of the Energy Loss Rate}

The first and simplest bound that may be obtained arises from 
assuming that all $\Aprime$ particles produced in the supernova core 
leave the supernova, carrying their energies with them. The constraint can be derived simply by 
requiring that the energy loss through this cooling channel be less than the cooling from neutrino emission; 
any greater, and it would have an observable effect on supernova cooling. This calculation yields values of 
$\varepsilon$ above which cooling through $\Aprime$ production is too rapid to be consistent with SN1987A. 
We will consider modifications to this bound from $\Aprime$ trapping and decay in subsequent subsections.

The quantity of interest is the rate of energy emission through dark gauge bosons. 
From the spin-summed, squared amplitudes described in the previous subsection, 
the energy emission rate is obtained by integrating over the phase space, 
and adding a factor of the energy of the emitted particle. To be specific, the 
energy emission rate per unit volume is 
\begin{widetext}
\beq
\label{eq:rate1}
Q_i = (2\pi)^4 \int\, E_{\mathrm{A}'} \, \sum_{s_1,s_2}\, \vert \mathcal{M}_i \vert^2 f(p_1) f(p_2)\delta^{(4)}(p_1+p_2-p_3-p_4-\qa)\, \dd \Pi,
\eeq
\end{widetext}
where 
\begin{equation}
d\Pi = \frac{\dd^3 \vec{q}_{\mathrm{A}'}}{(2\pi)^3 2E_{\mathrm{A}'}}\, \prod_{j=1}^{4}\, \frac{\dd^3 \vec{p}_j}{(2\pi^3) 2E_j}
\end{equation} 
is the Lorentz-invariant phase space interval, 
$ E_{\mathrm{A}'}$ is the energy of the emitted $\Aprime$ boson, 
$f(p)$ are the phase-space densities of the incoming nucleons, and 
the index $i$ on $Q_i$ refers to either the pp or pn processes. 
The nucleons in the core are comfortably non-degenerate and non-relativistic, 
so the Pauli blocking factor is omitted from Eq.~(\ref{eq:rate1}) and we take 
all nucleons to have a Maxwell-Boltzmann phase-space distribution distribution, 
\beq
f(p) =  \frac{n_{\mathrm{b}}}{2} \left( \frac{2 \pi}{M_{\mathrm{N}} T} \right)^{3/2} \exp \left(-\frac{p^2}{2 M_{\mathrm{N}} T} \right).
\eeq
We choose a baryon number density of $n_{\mathrm{b}} \approx 1.8 \times 10^{38}\, \mathrm{cm}^{-3}$ and a 
core supernova temperature of $T = 30\, \mathrm{MeV}$, both of which are typical choices and thought to 
be representative of supernova cores.

We performed the phase space integrals using the Monte Carlo routines in the {\tt CUBA} library \cite{Hahn:2004fe}.
We integrated over the momenta $\vec{p}_1$, $\vec{p}_2$, $\vec{p}_3$, and the direction of the 
three-momentum of the radiated boson $\hat{q}_{\mathrm{A}'}$, after fixing $\vec{p}_4$ 
and the magnitude of $\vec{q}_{\mathrm{A}'}$ using the delta functions.
 We used the {\tt suave} method provided within {\tt CUBA}, which combines 
importance sampling and adaptive subdivision, as this method provided the best compromise 
between accuracy and run-time for this particular application.

To obtain the dark gauge boson luminosity from $Q_{\mathrm{pp}}$ and $Q_{\mathrm{pn}}$, 
we assumed that $\Aprime$ production takes place in a stellar core of radius $\sim 1$~km, 
so that the total luminosity of $\Aprime$ is 
\begin{equation}
L_{\mathrm{A}'} = V (Q_{\mathrm{pp}} + Q_{\mathrm{pn}}),
\end{equation}
where $V$ is the volume of the sphere, and the luminosity per unit mass within the core is 
\begin{equation}
\left( \frac{L_{\mathrm{A'}}}{M} \right) = \frac{Q_{\mathrm{pp}} + Q_{\mathrm{pn}}}{\rho}, 
\end{equation}
where $\rho = 3 \times 10^{14}\, \mathrm{g/cm}^3$ is the mass density of the core.
Following previous studies, the energy loss rate into novel 
particles cannot exceed 
\begin{equation}
\epsilon_{\mathrm{A}'} = \left( \frac{L_{\mathrm{max}}}{M} \right) \approx 10^{19} \, \frac{\mathrm{erg}}{\mathrm{g}\cdot \mathrm{s}}
\end{equation}
without significantly reducing the duration of the neutrino burst observed at Earth for SN1987A 
\cite{raffelt96_book}. Writing $\epsilon_{\mathrm{A}'}  = \varepsilon^2 I_{\mathrm{A'}}(m_{\mathrm{A'}},T)$, 
we delineate the constraint on the dark photon mixing parameter by 
\begin{equation}
\label{eq:bound1}
\varepsilon \lesssim \sqrt{\frac{\epsilon_{\mathrm{A}'} }{I_{\mathrm{A}'}(m_{\mathrm{A}'}, T)}}.
\end{equation}
At this point, we note that Dent et al. in Ref.~\cite{dent_etal12} took a value of 
$\epsilon_{\mathrm{A}'} $ approximately three orders of magnitude larger than 
this generally accepted value. Remarkably, this nearly canceled the errors in their evaluation of 
$Q_{\mathrm{pp}}$ and $Q_{\mathrm{pn}}$, so that they quote constraints that are 
within an order of magnitude of the correct result.

The constraint derived in this manner from Eq.~(\ref{eq:bound1}) sets the lower limit on the 
exclusion band shown in Figure~\ref{fig:constraint}. We will discuss Fig.~\ref{fig:constraint} 
in more detail below. If all of the $\Aprime$ produced in the core leave the supernova freely, all 
values of $\varepsilon$ higher than those given by Eq.~(\ref{eq:bound1}) would be ruled out. 
However, as we have already mentioned, $\varepsilon$ can become sufficiently large that only a 
negligible amount of energy actually exits the supernova core. For large values or $\varepsilon$, 
this can occur because of either $\Aprime$ decays or $\Aprime$ scattering. These additional considerations place 
an upper limit on the values of $\varepsilon$ for which this constraint applies, 
and we discuss these effects in the next two subsections.

\subsection{The Decay Limit}

One upper limit to the excluded values of $\varepsilon$ may be found by considering 
decay of the dark bosons into Standard Model particles. Standard model particles will 
scatter and thermalize on a timescale much shorter than the timescale for the evolution 
of the core, so decays within the core contribute little to the supernova cooling. 

The dark boson has a typical lifetime of
\begin{equation}
\tau_{\mathrm{A}'} = \frac{3}{\varepsilon^2 \alpha m_{\mathrm{A}'}},
\end{equation}
where $\alpha$ is the fine structure constant. Therefore,
the typical travel distance to decay is given by 
\beq
\label{eq:decaylength}
l = \beta \tau_{\mathrm{A}'} = \frac{3 q_{\mathrm{A}'}}{\varepsilon^2 \alpha m_{\mathrm{A}'}^2},
\eeq
and so the fraction escaping the supernova before decaying may be estimated as (e.g., Ref.~\cite{bjorken_etal09})
\beq
\label{eq:decayfactor}
\exp \left( - \frac{r_{\mathrm{decay}}}{l} \right) = 
\exp \left( - \frac{r_{\mathrm{decay}} m_{\mathrm{A'}}^2\alpha \varepsilon^2}{3 q_{\mathrm{A'}}} \right),
\eeq
where $r_\mathrm{decay}$ is chosen to be 10 km, since the density of the supernova drops quickly around that size. This approximation should be valid so long as the size of the supernova within which standard model decay products can interact and be thermalized is significantly larger than the region within which $\Aprime$ are produced, an assumption that should be satisfied comfortably. To account for the suppression of gauge boson luminosity due to decays we simply multiply the phase space integrand in Eq.~(\ref{eq:rate1}) by the exponential suppression factor, after which the calculation proceeds as in the previous subsection. The limit is derived in the same way, with the additional complication that $I_{\mathrm{A}'}$ is now a function of $\varepsilon $, in addition to $m_{\mathrm{A}'}$ and $T$. The constraint Eq.~(\ref{eq:bound1}), therefore, becomes a transcendental equation that must be solved numerically. 

The luminosity in $\Aprime$ will be an increasing function of $\varepsilon$ until decays suppress the gauge boson luminosity, 
at which point $L_{\mathrm{A}'}$ becomes a rapidly decreasing function of $\varepsilon$. Therefore, the excluded values of 
$\varepsilon$ at a given mass will generally have a lower bound set by the calculations of the previous subsection, and an 
upper bound set by decays. An approximate treatment of the upper bound due to decays, as we present here, is 
sufficient because over almost the entire mass range of interest, larger values of $\varepsilon$ are independently 
excluded by terrestrial beam dump experiments \cite{bjorken_etal09}. Therefore, it is far more important to 
derive an accurate estimate of the {\em lower} boundary of the exclusion region (as was done in 
the previous subsection) than the upper boundary of the exclusion region.

\subsection{Trapping limit}

The second effect that produces an upper bound on the excluded region 
comes from considering trapping of dark bosons within the supernova. 
With a large enough coupling, the dark photons will thermalize and then 
will be emitted from an approximately spherical ``dark photosphere" at the radial position 
where the $\Aprime$ mean free path becomes larger than the typical size of the supernova core. 
In this case the luminosity is given simply by Stefan's law, 
\beq
L_{\mathrm{A'}}  = 4\pi \rdp^2 T_{\mathrm{A'}}^4 \sigma,
\eeq
where $\rdp$ is now the radius of the emitting shell and $T_{\mathrm{A'}}$ its temperature. 
We estimate the radius of this dark photosphere as $\rdp=10$ km, 
because the density of the supernova drops drastically around that point. We will 
confirm shortly that this estimate is consistent within the context of the simple model that we 
adopt for the interior structure of the supernova atmospherer. The bound on the luminosity can then 
be recast as a bound on $T_{\mathrm{A}'}$, 
\beq
T_{\mathrm{A}'} \lesssim 9.6\rm\ MeV.
\eeq
That bound can then be translated into the desired bound on the coupling as a function of mass by adopting a 
simple model for the supernova atmosphere, assuming that the particles are emitted from the dark photosphere 
at a point where the optical depth to scattering is $\tau = 2/3 $, and finding the temperature that corresponds 
to that optical depth. 

This is a somewhat involved calculation. First, one needs a model for the density and temperature in the supernova. Following the simple, early model of Ref.~\cite{turner88}, we assume a simple power-law model for the supernova core, with 
\bea
\rho &=& \rho_{\mathrm{p}} \left( \frac{R}{r}\right )^n,\\
T &=& T_{\mathrm{R}} \left[ \frac{\rho(r)}{\rho_{\mathrm{p}}} \right]^{1/3},
\eea
with $  \rho_{\mathrm{p}}  = 3\times10^{14}\ {\rm g/cm}^3$, $T_{\mathrm{R}} = 30$~MeV, $R = 10$ km,
and $n = 5$.
The optical depth is given by 
\beq
\tau = \int_{r_x}^{\infty}\, \kappa \rho\,  \dd r,
\eeq
where $ \kappa$ is the opacity, which we take to be the Rosseland mean 
\beq 
\frac{1}{\kappa \rho} = \frac{15}{4 \pi^4 T^5}\, \int_{M_{\mathrm{A'}}}^{\infty}\, 
\frac{E_{\mathrm{A'}}^2\, e^{E_{\mathrm{A'}}/T} 
\sqrt{E_{\mathrm{A'}}^2 - m_{\mathrm{A'}}^2}}
{(e^{E_{\mathrm{A'}}/T}-1)^2}\, l_{\mathrm{A'}}\, \dd E_{\mathrm{A'}},
\eeq
where $l_{\mathrm{A'}}$ is the mean free path.

The inverse mean free path can be obtained by modifying $ Q_i $, 
the expression for the energy loss rate, as follows: removing the factor of $ E_\mathrm{A'} $ 
and the phase space integral over $ \qa $, and adding a factor of $ e^{E_\mathrm{A'}/T} $ for detailed balance. This factor comes from turning the $\mathrm{A'}$ from an outgoing to an incoming state in the calculation (e.g., see Eq.~(4.43) in Ref.~\cite{raffelt96_book}).
This gives the inverse mean free path as a function of mass and coupling. 
Again, we perform the required integration numerically using the {\tt CUBA }package. 
This then allows the calculation of $\kappa_{\mathrm{pp}}$ and $\kappa_{\mathrm{pn}}$, 
which are the opacities due to inverse bremsstrahlung for proton-proton and proton-neutron 
processes respectively. These dominate the opacity of the star to $\mathrm{A'}$ propagation. 
The inverse opacities for the $\mathrm{pn}$ and $\mathrm{pp}$ processes add, 
giving the total opacity $ \kappa^{-1} = \kappa_{\mathrm{pp}}^{-1} + \kappa_{\mathrm{pn}}^{-1} $

Having obtained an expression for $ \kappa $, we can now find the optical depth as follows. 
Define a typical optical depth as $\tau_\mathrm{R} = \kappa_\mathrm{R} \rho_\mathrm{R} R $. We then have 
\beq 
\kappa \rho R = \tau_\mathrm{R} \left( \frac{\rho}{\rho_\mathrm{R}} \right)^2 \left( \frac{T_\mathrm{R}}{T} \right)^{3/2}.
\eeq
This can be combined with the previous expressions for the density and temperature as a function of position 
and plugged in to the integral expression for the optical depth to obtain 
\bea 
\tau &=& \int_{r_x}^{\infty}\, \tau_\mathrm{R} \left( \frac{R}{r} \right)^{3n/2} \dd r \\
 &=& \frac{\tau_\mathrm{r}}{\frac{3n}{2}-1} \left( \frac{T_\mathrm{A'}}{T_\mathrm{R}} \right)^{(9/2-3/n)}. 
 \eea
 The bound on the coupling is obtained by requiring  $ \tau(\varepsilon,m_\mathrm{A'}) \le 2/3 $. Note that strictly speaking we should have determined $ r_{\mathrm{dp}}$ rather than assuming a value. However, it is possible to verify the self-consistency of our assumption using this model for the supernova atmosphere. In particular, this model implies that an optical depth of $\tau=2/3$ is reached at a radial position of $ r_{\mathrm{dp}}= 11$ km,  validating the assumption made at the outset. As we stated in the previous subsection, the approximate treatment of the upper limit of the exclusion region is justified by the fact that values of $\varepsilon$ close to the upper limit of the exclusion region are ruled out by independent, terrestrial experiments.

\section{Results}
\label{section:results}

\begin{figure*}[th]
\includegraphics[width=14cm]{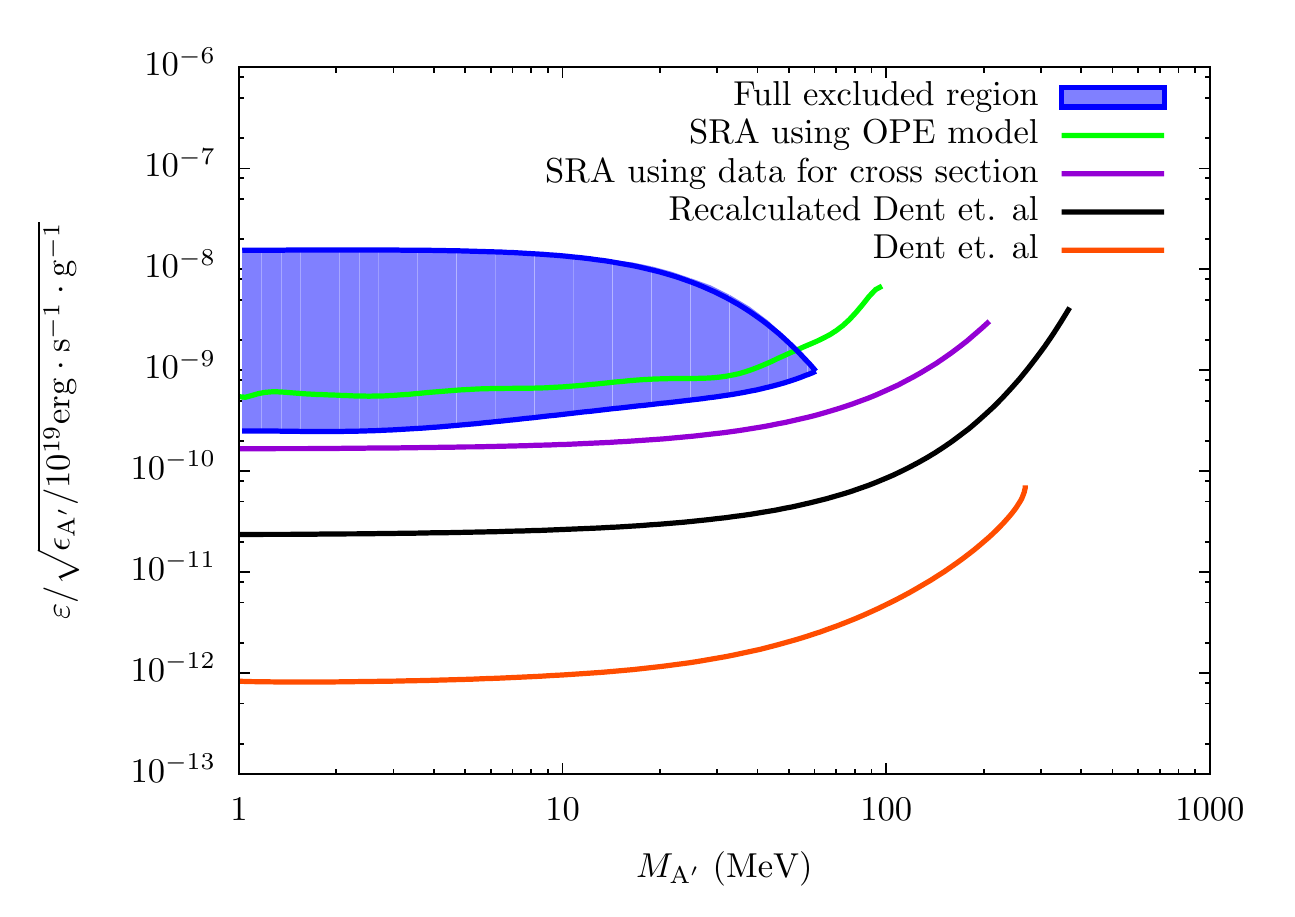}
\caption{
Constraints on dark photon models. The constraint scales according to the assumed, maximal energy loss rate of $\epsilon_{\mathrm{A}'} = 10^{19}\, \mathrm{erg}\, \mathrm{g}^{-1}\ \mathrm{s}^{-1}$ as shown on the vertical axis label. Our primary result is the excluded region 
in the $\varepsilon$-$m_{\mathrm{A'}}$ plane shaded in blue. The black curve is our attempt to reproduce exactly Ref.~\cite{dent_etal12}, while the red curve is Ref.~\cite{dent_etal12} but with the correct luminosity constraint.  The purple curve was computed by using the SRA and fixing the cross section through a polynomial fit to nuclear scattering data (e.g., see \cite{rrapaj_reddy16}), while the green curve uses the SRA of the OPE result.
}
\label{fig:constraint}
\end{figure*}

The considerations of the previous section lead to constraints on the 
viable parameters for dark electromagnetism models of SIDM, in particular, 
on the mixing parameter $\varepsilon$ as a function of the dark photon mass $m_{\mathrm{A'}}$. 
The excluded region is depicted in Fig.~\ref{fig:constraint} as the shaded blue region. The 
lower limit of our excluded region lies at a value of $\varepsilon$ about a factor of four 
larger than that of Dent et al. in Ref.~\cite{dent_etal12} and is in excellent agreement 
with the result of Rrapaj and Reddy \cite{rrapaj_reddy16} despite the fact that they work 
with an SRA approximation and we work with a OPE model.

Figure~\ref{fig:constraint} shows several other estimates of the lower bound on 
$\varepsilon$ in an effort to elucidate possibly confusing points in the existing 
literature. As we stated earlier, the work of Ref.~\cite{dent_etal12} contained several 
errors that partially cancelled each other. The black line in Fig.~\ref{fig:constraint} represents 
our effort to reproduce the result of Ref.~\cite{dent_etal12} using (incorrectly) the equations 
that they quote in their paper. However, Ref.~\cite{dent_etal12} had significant errors in 
their calculations of the dark photon luminosity and chose an energy loss rate three orders 
of magnitude higher than other practitioners. Adopting the Dent et al. \cite{dent_etal12} 
equations for the dark photon luminosity, but the correct limit on the energy loss rate, 
results in the red line in Fig.~\ref{fig:constraint}. As Ref.~\cite{dent_etal12}, our primary 
result uses a OPE model for nucleon interactions, so the entirety of the difference between 
the red line and the lower bound of our exclusion region stems from an incorrect estimate 
of the bremsstrahlung rate in Ref.~\cite{dent_etal12}.

Rrapaj and Reddy \cite{rrapaj_reddy16} estimated $\mathrm{A'}$ luminosity 
using the SRA for nucleon interactions, which enabled them to use 
nuclear scattering data in the estimation of the bremsstrahlung rate. 
This approach overcomes the shortcomings of the OPE approximation, 
though at high dark photon masses this approximation breaks down. 
We have repeated the calculation in the SRA as described in 
Rrapaj and Reddy \cite{rrapaj_reddy16} and our result is 
shown as the purple line in Fig.~\ref{fig:constraint}. 
Our SRA result is in excellent agreement with Ref.~\cite{rrapaj_reddy16}. It is 
also worth noting that the SRA and OPE approximations yield constraints 
on $\varepsilon$ that agree quite well with each other. This suggests, of course, 
that the discrepancy between Ref.~\cite{dent_etal12} and Ref.~\cite{rrapaj_reddy16} 
stems primarily from errors in Ref.~\cite{dent_etal12} and not to the OPE model 
of nucleon interactions. In an effort to estimate 
the possible shortcomings of the SRA approximation, the green line in 
Fig.~\ref{fig:constraint} shows the lower bound on the excluded region that 
we drive using the SRA along with OPE expressions for nucleon scattering 
cross sections. As one can see, the SRA is in good agreement with the full 
OPE bound. Consequently, we suggest that use of the SRA in this context 
results in an overestimate of the lower limit on $\varepsilon$ of less than 
a factor or three. 

\begin{figure*}[th]
	\includegraphics[width=14cm]{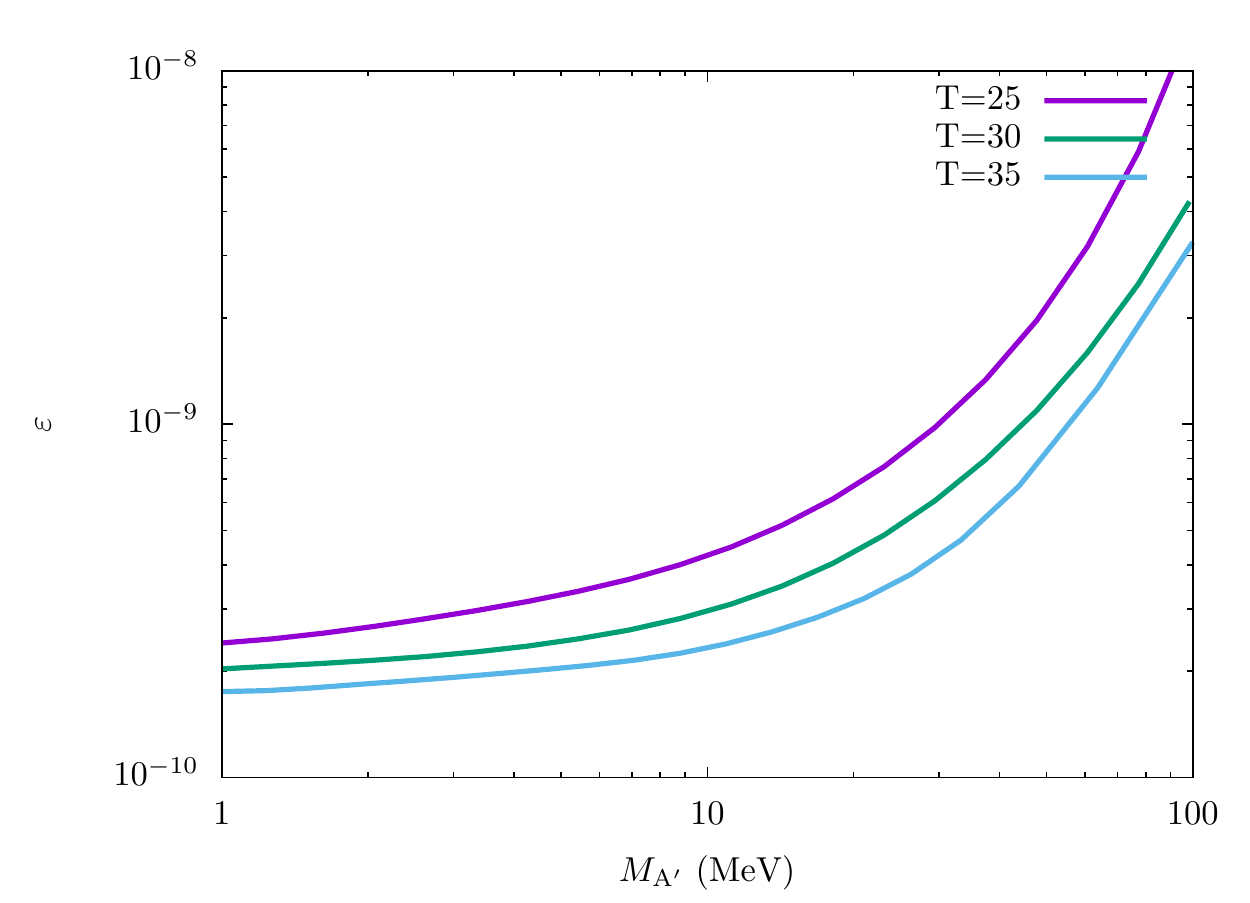}
	\caption{Effect of changing the core temperature on the lower bound of the dark photon constraint. The  calculation that produced the blue line of Fig. \ref{fig:constraint} was repeated for core temperatures of 25 MeV, 30 MeV, and 35 MeV. In this case the energy loss rate was throughout taken to be $\epsilon_{\mathrm{A}'} = 10^{19}\, \mathrm{erg}\, \mathrm{g}^{-1}\ \mathrm{s}^{-1}$
	}
	\label{fig:temperature}
\end{figure*}

\section{Discussion}
\label{section:conclusions}

We have revisited constraints on SIDM models in which the 
self-interaction arises from dark electromagnetism. We have 
constrained the mixing parameter for models in which the dark photon 
mixes with the standard model photon through a kinetic mixing term. Our calculation 
is similar to previous work in Ref.~\cite{dent_etal12} and Ref.~\cite{rrapaj_reddy16} and 
is aimed at clarifying some confusion in the literature on this subject that may stem 
from several errors in the calculation of Ref.~\cite{dent_etal12}.

Our constraints are on the mixing parameter are shown in 
Fig.~\ref{fig:constraint} and agree well with the those presented in 
Ref.~\cite{rrapaj_reddy16} and, for dark photon masses $m_{\mathrm{A}'} \gtrsim 10$~MeV, 
Ref.~\cite{hardy_lasenby17}. We argue that the lower bound on the dark photon mixing parameter 
is only sensitive to the SRA to within a factor of $\sim 3$ or less. A more precise constraint 
on such models from SN1987A is probably not practicable because the interior temperature 
of the SN1987A core is somewhat uncertain and this uncertainty alone gives rise to an uncertainty 
in the excluded region of at least a factor of $\sim 2$. Finally, \citet{hardy_lasenby17} consider 
plasma effects within the supernova. The most important ramification of this work in the context of 
our paper is that it demonstrates that constraints on $\varepsilon$ should be significantly weakened 
relative to our results in the mass range $m_{\mathrm{A}'} \lesssim 10$~MeV.

Finally, we note that the interior temperature and density within the progenitor of SN1987A remain 
uncertain at least at the level of several tens of percent \citep[e.g.][]{perego15,farmer16}. This is well worth 
noting because the constraints one derives using the techniques in this work or similar techniques 
are quite sensitive to the properties of the SN1987A core, particularly the core temperature. 
As an example of this, we have repeated a portion of our calculations after 
shifting the core temperature by $\pm 5\, \mathrm{MeV}$. This result is shown in Figure~\ref{fig:temperature}.  
We find that the lower boundary of the excluded region shifts by a factor of $\sim 2$ at low 
mass ($m_{\mathrm{A}'} \lesssim T$ where $T = 30\, \mathrm{MeV}$ in our fiducial calculation, following 
work of previous authors) and considerably more at high mass. We conclude that the differences among theoretical 
techniques for computing the excluded region (OPE, SRA, ...) lead to uncertainties that are of the same order or 
smaller than the uncertainty in the excluded region induced by our limited knowledge of the 
properties of the SN1987A progenitor interior. The excluded regions we and other authors delineate 
should be considered to be uncertain by a factor of $\sim 3$. Refining the theoretical approach to this 
problem is therefore unwarranted until a breakthrough in our understanding of the interior of the 
SN1987A progenitor is achieved.

\acknowledgments

We thank Carles Badenes, John Hillier, and Francis Timmes for guidance in understanding 
the state of knowledge of the SN1987A progenitor. 
AKL was supported in part by NSF grant PHY-1519175. 
ARZ was supported in part by the DoE through Grant DE-SC0007914 and by 
the Pittsburgh Particle physics Astrophysics and Cosmology Center (Pitt PACC) 
at the University of Pittsburgh.

\bibliography{draft}

\begin{thebibliography}{62}
\expandafter\ifx\csname natexlab\endcsname\relax\def\natexlab#1{#1}\fi
\expandafter\ifx\csname bibnamefont\endcsname\relax
  \def\bibnamefont#1{#1}\fi
\expandafter\ifx\csname bibfnamefont\endcsname\relax
  \def\bibfnamefont#1{#1}\fi
\expandafter\ifx\csname citenamefont\endcsname\relax
  \def\citenamefont#1{#1}\fi
\expandafter\ifx\csname url\endcsname\relax
  \def\url#1{\texttt{#1}}\fi
\expandafter\ifx\csname urlprefix\endcsname\relax\def\urlprefix{URL }\fi
\providecommand{\bibinfo}[2]{#2}
\providecommand{\eprint}[2][]{\url{#2}}

\bibitem[{\citenamefont{{Jungman} et~al.}(1996)\citenamefont{{Jungman},
  {Kamionkowski}, and {Griest}}}]{jungman_etal96}
\bibinfo{author}{\bibfnamefont{G.}~\bibnamefont{{Jungman}}},
  \bibinfo{author}{\bibfnamefont{M.}~\bibnamefont{{Kamionkowski}}},
  \bibnamefont{and} \bibinfo{author}{\bibfnamefont{K.}~\bibnamefont{{Griest}}},
  \bibinfo{journal}{\physrep} \textbf{\bibinfo{volume}{267}},
  \bibinfo{pages}{195} (\bibinfo{year}{1996}).

\bibitem[{\citenamefont{{Carlson} et~al.}(1992)\citenamefont{{Carlson},
  {Machacek}, and {Hall}}}]{carlson_etal92}
\bibinfo{author}{\bibfnamefont{E.~D.} \bibnamefont{{Carlson}}},
  \bibinfo{author}{\bibfnamefont{M.~E.} \bibnamefont{{Machacek}}},
  \bibnamefont{and} \bibinfo{author}{\bibfnamefont{L.~J.}
  \bibnamefont{{Hall}}}, \bibinfo{journal}{\apj}
  \textbf{\bibinfo{volume}{398}}, \bibinfo{pages}{43} (\bibinfo{year}{1992}).

\bibitem[{\citenamefont{{de Laix} et~al.}(1995)\citenamefont{{de Laix},
  {Scherrer}, and {Schaefer}}}]{deLaix_etal95}
\bibinfo{author}{\bibfnamefont{A.~A.} \bibnamefont{{de Laix}}},
  \bibinfo{author}{\bibfnamefont{R.~J.} \bibnamefont{{Scherrer}}},
  \bibnamefont{and} \bibinfo{author}{\bibfnamefont{R.~K.}
  \bibnamefont{{Schaefer}}}, \bibinfo{journal}{\apj}
  \textbf{\bibinfo{volume}{452}}, \bibinfo{pages}{495} (\bibinfo{year}{1995}),
  \eprint{arXiv:astro-ph/9502087}.

\bibitem[{\citenamefont{{Atrio-Barandela} and
  {Davidson}}(1997)}]{atrio-barandela_davidson97}
\bibinfo{author}{\bibfnamefont{F.}~\bibnamefont{{Atrio-Barandela}}}
  \bibnamefont{and}
  \bibinfo{author}{\bibfnamefont{S.}~\bibnamefont{{Davidson}}},
  \bibinfo{journal}{\prd} \textbf{\bibinfo{volume}{55}}, \bibinfo{pages}{5886}
  (\bibinfo{year}{1997}), \eprint{arXiv:astro-ph/9702236}.

\bibitem[{\citenamefont{{Spergel} and
  {Steinhardt}}(2000)}]{spergel_steinhardt00}
\bibinfo{author}{\bibfnamefont{D.~N.} \bibnamefont{{Spergel}}}
  \bibnamefont{and} \bibinfo{author}{\bibfnamefont{P.~J.}
  \bibnamefont{{Steinhardt}}}, \bibinfo{journal}{Phys. Rev. Lett.}
  \textbf{\bibinfo{volume}{84}}, \bibinfo{pages}{3760} (\bibinfo{year}{2000}),
  \eprint{arXiv:astro-ph/9909386}.

\bibitem[{\citenamefont{{Hogan} and {Dalcanton}}(2000)}]{hogan_dalcanton00}
\bibinfo{author}{\bibfnamefont{C.~J.} \bibnamefont{{Hogan}}} \bibnamefont{and}
  \bibinfo{author}{\bibfnamefont{J.~J.} \bibnamefont{{Dalcanton}}},
  \bibinfo{journal}{\prd} \textbf{\bibinfo{volume}{62}},
  \bibinfo{pages}{063511} (\bibinfo{year}{2000}),
  \eprint{arXiv:astro-ph/0002330}.

\bibitem[{\citenamefont{{Mohapatra} and {Teplitz}}(2000)}]{mohapatra_teplitz00}
\bibinfo{author}{\bibfnamefont{R.~N.} \bibnamefont{{Mohapatra}}}
  \bibnamefont{and} \bibinfo{author}{\bibfnamefont{V.~L.}
  \bibnamefont{{Teplitz}}}, \bibinfo{journal}{\prd}
  \textbf{\bibinfo{volume}{62}}, \bibinfo{pages}{063506}
  (\bibinfo{year}{2000}), \eprint{arXiv:astro-ph/0001362}.

\bibitem[{\citenamefont{{Dav{\'e}} et~al.}(2001)\citenamefont{{Dav{\'e}},
  {Spergel}, {Steinhardt}, and {Wandelt}}}]{dave_etal01}
\bibinfo{author}{\bibfnamefont{R.}~\bibnamefont{{Dav{\'e}}}},
  \bibinfo{author}{\bibfnamefont{D.~N.} \bibnamefont{{Spergel}}},
  \bibinfo{author}{\bibfnamefont{P.~J.} \bibnamefont{{Steinhardt}}},
  \bibnamefont{and} \bibinfo{author}{\bibfnamefont{B.~D.}
  \bibnamefont{{Wandelt}}}, \bibinfo{journal}{\apj}
  \textbf{\bibinfo{volume}{547}}, \bibinfo{pages}{574} (\bibinfo{year}{2001}),
  \eprint{arXiv:astro-ph/0006218}.

\bibitem[{\citenamefont{{Hisano} et~al.}(2004)\citenamefont{{Hisano},
  {Matsumoto}, and {Nojiri}}}]{hisano_etal04}
\bibinfo{author}{\bibfnamefont{J.}~\bibnamefont{{Hisano}}},
  \bibinfo{author}{\bibfnamefont{S.}~\bibnamefont{{Matsumoto}}},
  \bibnamefont{and} \bibinfo{author}{\bibfnamefont{M.~M.}
  \bibnamefont{{Nojiri}}}, \bibinfo{journal}{Phys. Rev. Lett.}
  \textbf{\bibinfo{volume}{92}}, \bibinfo{pages}{031303}
  (\bibinfo{year}{2004}), \eprint{arXiv:hep-ph/0307216}.

\bibitem[{\citenamefont{{Hisano} et~al.}(2005)\citenamefont{{Hisano},
  {Matsumoto}, {Nojiri}, and {Saito}}}]{hisano_etal05}
\bibinfo{author}{\bibfnamefont{J.}~\bibnamefont{{Hisano}}},
  \bibinfo{author}{\bibfnamefont{S.}~\bibnamefont{{Matsumoto}}},
  \bibinfo{author}{\bibfnamefont{M.~M.} \bibnamefont{{Nojiri}}},
  \bibnamefont{and} \bibinfo{author}{\bibfnamefont{O.}~\bibnamefont{{Saito}}},
  \bibinfo{journal}{\prd} \textbf{\bibinfo{volume}{71}},
  \bibinfo{pages}{063528} (\bibinfo{year}{2005}),
  \eprint{arXiv:hep-ph/0412403}.

\bibitem[{\citenamefont{{Pospelov} et~al.}(2008)\citenamefont{{Pospelov},
  {Ritz}, and {Voloshin}}}]{pospelov_etal08}
\bibinfo{author}{\bibfnamefont{M.}~\bibnamefont{{Pospelov}}},
  \bibinfo{author}{\bibfnamefont{A.}~\bibnamefont{{Ritz}}}, \bibnamefont{and}
  \bibinfo{author}{\bibfnamefont{M.}~\bibnamefont{{Voloshin}}},
  \bibinfo{journal}{Phys. Lett. B} \textbf{\bibinfo{volume}{662}},
  \bibinfo{pages}{53} (\bibinfo{year}{2008}), \eprint{0711.4866}.

\bibitem[{\citenamefont{{Arkani-Hamed}
  et~al.}(2008)\citenamefont{{Arkani-Hamed}, {Finkbeiner}, {Slatyer}, and
  {Weiner}}}]{arkani-hamed_etal08a}
\bibinfo{author}{\bibfnamefont{N.}~\bibnamefont{{Arkani-Hamed}}},
  \bibinfo{author}{\bibfnamefont{D.~P.} \bibnamefont{{Finkbeiner}}},
  \bibinfo{author}{\bibfnamefont{T.~R.} \bibnamefont{{Slatyer}}},
  \bibnamefont{and} \bibinfo{author}{\bibfnamefont{N.}~\bibnamefont{{Weiner}}},
  \bibinfo{journal}{ArXiv e-prints}  (\bibinfo{year}{2008}),
  \eprint{0810.0713}.

\bibitem[{\citenamefont{{Lattanzi} and {Silk}}(2008)}]{lattanzi_silk08}
\bibinfo{author}{\bibfnamefont{M.}~\bibnamefont{{Lattanzi}}} \bibnamefont{and}
  \bibinfo{author}{\bibfnamefont{J.}~\bibnamefont{{Silk}}},
  \bibinfo{journal}{ArXiv e-prints}  (\bibinfo{year}{2008}),
  \eprint{0812.0360}.

\bibitem[{\citenamefont{{Ackerman} et~al.}(2009)\citenamefont{{Ackerman},
  {Buckley}, {Carroll}, and {Kamionkowski}}}]{ackerman_etal09}
\bibinfo{author}{\bibfnamefont{L.}~\bibnamefont{{Ackerman}}},
  \bibinfo{author}{\bibfnamefont{M.~R.} \bibnamefont{{Buckley}}},
  \bibinfo{author}{\bibfnamefont{S.~M.} \bibnamefont{{Carroll}}},
  \bibnamefont{and}
  \bibinfo{author}{\bibfnamefont{M.}~\bibnamefont{{Kamionkowski}}},
  \bibinfo{journal}{\prd} \textbf{\bibinfo{volume}{79}},
  \bibinfo{pages}{023519} (\bibinfo{year}{2009}), \eprint{0810.5126}.

\bibitem[{\citenamefont{{Feng} et~al.}(2009)\citenamefont{{Feng}, {Kaplinghat},
  {Tu}, and {Yu}}}]{feng_etal09}
\bibinfo{author}{\bibfnamefont{J.~L.} \bibnamefont{{Feng}}},
  \bibinfo{author}{\bibfnamefont{M.}~\bibnamefont{{Kaplinghat}}},
  \bibinfo{author}{\bibfnamefont{H.}~\bibnamefont{{Tu}}}, \bibnamefont{and}
  \bibinfo{author}{\bibfnamefont{H.-B.} \bibnamefont{{Yu}}},
  \bibinfo{journal}{Journal of Cosmology and Astro-Particle Physics}
  \textbf{\bibinfo{volume}{7}}, \bibinfo{pages}{4} (\bibinfo{year}{2009}),
  \eprint{0905.3039}.

\bibitem[{\citenamefont{{Kong} et~al.}(2015)\citenamefont{{Kong}, {Mohlabeng},
  and {Park}}}]{kong_etal15}
\bibinfo{author}{\bibfnamefont{K.}~\bibnamefont{{Kong}}},
  \bibinfo{author}{\bibfnamefont{G.}~\bibnamefont{{Mohlabeng}}},
  \bibnamefont{and} \bibinfo{author}{\bibfnamefont{J.-C.}
  \bibnamefont{{Park}}}, \bibinfo{journal}{Physics Letters B}
  \textbf{\bibinfo{volume}{743}}, \bibinfo{pages}{256} (\bibinfo{year}{2015}),
  \eprint{1411.6632}.

\bibitem[{\citenamefont{Buckley and Fox}(2010)}]{Buckley:2009in}
\bibinfo{author}{\bibfnamefont{M.~R.} \bibnamefont{Buckley}} \bibnamefont{and}
  \bibinfo{author}{\bibfnamefont{P.~J.} \bibnamefont{Fox}},
  \bibinfo{journal}{Phys. Rev.} \textbf{\bibinfo{volume}{D81}},
  \bibinfo{pages}{083522} (\bibinfo{year}{2010}), \eprint{0911.3898}.

\bibitem[{\citenamefont{Boddy et~al.}(2014{\natexlab{a}})\citenamefont{Boddy,
  Feng, Kaplinghat, and Tait}}]{Boddy:2014yra}
\bibinfo{author}{\bibfnamefont{K.~K.} \bibnamefont{Boddy}},
  \bibinfo{author}{\bibfnamefont{J.~L.} \bibnamefont{Feng}},
  \bibinfo{author}{\bibfnamefont{M.}~\bibnamefont{Kaplinghat}},
  \bibnamefont{and} \bibinfo{author}{\bibfnamefont{T.~M.~P.}
  \bibnamefont{Tait}}, \bibinfo{journal}{Phys. Rev.}
  \textbf{\bibinfo{volume}{D89}}, \bibinfo{pages}{115017}
  (\bibinfo{year}{2014}{\natexlab{a}}), \eprint{1402.3629}.

\bibitem[{\citenamefont{Boddy et~al.}(2014{\natexlab{b}})\citenamefont{Boddy,
  Feng, Kaplinghat, Shadmi, and Tait}}]{Boddy:2014qxa}
\bibinfo{author}{\bibfnamefont{K.~K.} \bibnamefont{Boddy}},
  \bibinfo{author}{\bibfnamefont{J.~L.} \bibnamefont{Feng}},
  \bibinfo{author}{\bibfnamefont{M.}~\bibnamefont{Kaplinghat}},
  \bibinfo{author}{\bibfnamefont{Y.}~\bibnamefont{Shadmi}}, \bibnamefont{and}
  \bibinfo{author}{\bibfnamefont{T.~M.~P.} \bibnamefont{Tait}},
  \bibinfo{journal}{Phys. Rev.} \textbf{\bibinfo{volume}{D90}},
  \bibinfo{pages}{095016} (\bibinfo{year}{2014}{\natexlab{b}}),
  \eprint{1408.6532}.

\bibitem[{\citenamefont{Boddy et~al.}(2016)\citenamefont{Boddy, Kaplinghat,
  Kwa, and Peter}}]{Boddy:2016bbu}
\bibinfo{author}{\bibfnamefont{K.~K.} \bibnamefont{Boddy}},
  \bibinfo{author}{\bibfnamefont{M.}~\bibnamefont{Kaplinghat}},
  \bibinfo{author}{\bibfnamefont{A.}~\bibnamefont{Kwa}}, \bibnamefont{and}
  \bibinfo{author}{\bibfnamefont{A.~H.~G.} \bibnamefont{Peter}},
  \bibinfo{journal}{Phys. Rev.} \textbf{\bibinfo{volume}{D94}},
  \bibinfo{pages}{123017} (\bibinfo{year}{2016}), \eprint{1609.03592}.

\bibitem[{\citenamefont{{Yoshida} et~al.}(2000)\citenamefont{{Yoshida},
  {Springel}, {White}, and {Tormen}}}]{yoshida_etal00}
\bibinfo{author}{\bibfnamefont{N.}~\bibnamefont{{Yoshida}}},
  \bibinfo{author}{\bibfnamefont{V.}~\bibnamefont{{Springel}}},
  \bibinfo{author}{\bibfnamefont{S.~D.~M.} \bibnamefont{{White}}},
  \bibnamefont{and} \bibinfo{author}{\bibfnamefont{G.}~\bibnamefont{{Tormen}}},
  \bibinfo{journal}{\apjl} \textbf{\bibinfo{volume}{544}}, \bibinfo{pages}{L87}
  (\bibinfo{year}{2000}), \eprint{arXiv:astro-ph/0006134}.

\bibitem[{\citenamefont{{Gnedin} and {Ostriker}}(2001)}]{gnedin_ostriker01}
\bibinfo{author}{\bibfnamefont{O.~Y.} \bibnamefont{{Gnedin}}} \bibnamefont{and}
  \bibinfo{author}{\bibfnamefont{J.~P.} \bibnamefont{{Ostriker}}},
  \bibinfo{journal}{\apj} \textbf{\bibinfo{volume}{561}}, \bibinfo{pages}{61}
  (\bibinfo{year}{2001}), \eprint{arXiv:astro-ph/0010436}.

\bibitem[{\citenamefont{{Miralda-Escud{\'e}}}(2002)}]{miralda-escude02}
\bibinfo{author}{\bibfnamefont{J.}~\bibnamefont{{Miralda-Escud{\'e}}}},
  \bibinfo{journal}{\apj} \textbf{\bibinfo{volume}{564}}, \bibinfo{pages}{60}
  (\bibinfo{year}{2002}).

\bibitem[{\citenamefont{{Randall} et~al.}(2008)\citenamefont{{Randall},
  {Markevitch}, {Clowe}, {Gonzalez}, and {Brada{\v c}}}}]{randall_etal08}
\bibinfo{author}{\bibfnamefont{S.~W.} \bibnamefont{{Randall}}},
  \bibinfo{author}{\bibfnamefont{M.}~\bibnamefont{{Markevitch}}},
  \bibinfo{author}{\bibfnamefont{D.}~\bibnamefont{{Clowe}}},
  \bibinfo{author}{\bibfnamefont{A.~H.} \bibnamefont{{Gonzalez}}},
  \bibnamefont{and} \bibinfo{author}{\bibfnamefont{M.}~\bibnamefont{{Brada{\v
  c}}}}, \bibinfo{journal}{\apj} \textbf{\bibinfo{volume}{679}},
  \bibinfo{pages}{1173} (\bibinfo{year}{2008}), \eprint{0704.0261}.

\bibitem[{\citenamefont{{Kamionkowski} and
  {Profumo}}(2008)}]{kamionkowski_profumo08}
\bibinfo{author}{\bibfnamefont{M.}~\bibnamefont{{Kamionkowski}}}
  \bibnamefont{and}
  \bibinfo{author}{\bibfnamefont{S.}~\bibnamefont{{Profumo}}},
  \bibinfo{journal}{Phys. Rev. Lett.} \textbf{\bibinfo{volume}{101}},
  \bibinfo{pages}{261301} (\bibinfo{year}{2008}).

\bibitem[{\citenamefont{{Zentner}}(2009)}]{zentner09}
\bibinfo{author}{\bibfnamefont{A.~R.} \bibnamefont{{Zentner}}},
  \bibinfo{journal}{\prd} \textbf{\bibinfo{volume}{80}}, \bibinfo{eid}{063501}
  (\bibinfo{year}{2009}), \eprint{0907.3448}.

\bibitem[{\citenamefont{{Robertson} and {Zentner}}(2009)}]{robertson_zentner09}
\bibinfo{author}{\bibfnamefont{B.~E.} \bibnamefont{{Robertson}}}
  \bibnamefont{and} \bibinfo{author}{\bibfnamefont{A.~R.}
  \bibnamefont{{Zentner}}}, \bibinfo{journal}{\prd}
  \textbf{\bibinfo{volume}{79}}, \bibinfo{pages}{083525}
  (\bibinfo{year}{2009}), \eprint{0902.0362}.

\bibitem[{\citenamefont{{Pieri} et~al.}(2009)\citenamefont{{Pieri}, {Lattanzi},
  and {Silk}}}]{pieri_etal09}
\bibinfo{author}{\bibfnamefont{L.}~\bibnamefont{{Pieri}}},
  \bibinfo{author}{\bibfnamefont{M.}~\bibnamefont{{Lattanzi}}},
  \bibnamefont{and} \bibinfo{author}{\bibfnamefont{J.}~\bibnamefont{{Silk}}},
  \bibinfo{journal}{ArXiv e-prints}  (\bibinfo{year}{2009}),
  \eprint{0902.4330}.

\bibitem[{\citenamefont{{Spolyar} et~al.}(2009)\citenamefont{{Spolyar},
  {Buckley}, {Freese}, {Hooper}, and {Murayama}}}]{spolyar_etal09}
\bibinfo{author}{\bibfnamefont{D.}~\bibnamefont{{Spolyar}}},
  \bibinfo{author}{\bibfnamefont{M.}~\bibnamefont{{Buckley}}},
  \bibinfo{author}{\bibfnamefont{K.}~\bibnamefont{{Freese}}},
  \bibinfo{author}{\bibfnamefont{D.}~\bibnamefont{{Hooper}}}, \bibnamefont{and}
  \bibinfo{author}{\bibfnamefont{H.}~\bibnamefont{{Murayama}}},
  \bibinfo{journal}{ArXiv e-prints}  (\bibinfo{year}{2009}),
  \eprint{0905.4764}.

\bibitem[{\citenamefont{{Finkbeiner} et~al.}(2009)\citenamefont{{Finkbeiner},
  {Lin}, and {Weiner}}}]{finkbeiner_etal09}
\bibinfo{author}{\bibfnamefont{D.~P.} \bibnamefont{{Finkbeiner}}},
  \bibinfo{author}{\bibfnamefont{T.}~\bibnamefont{{Lin}}}, \bibnamefont{and}
  \bibinfo{author}{\bibfnamefont{N.}~\bibnamefont{{Weiner}}},
  \bibinfo{journal}{ArXiv e-prints}  (\bibinfo{year}{2009}),
  \eprint{0906.0002}.

\bibitem[{\citenamefont{{Slatyer} et~al.}(2009)\citenamefont{{Slatyer},
  {Padmanabhan}, and {Finkbeiner}}}]{slatyer_etal09}
\bibinfo{author}{\bibfnamefont{T.~R.} \bibnamefont{{Slatyer}}},
  \bibinfo{author}{\bibfnamefont{N.}~\bibnamefont{{Padmanabhan}}},
  \bibnamefont{and} \bibinfo{author}{\bibfnamefont{D.~P.}
  \bibnamefont{{Finkbeiner}}}, \bibinfo{journal}{ArXiv e-prints}
  (\bibinfo{year}{2009}), \eprint{0906.1197}.

\bibitem[{\citenamefont{{Bramante} et~al.}(2014)\citenamefont{{Bramante},
  {Fukushima}, {Kumar}, and {Stopnitzky}}}]{bramante_etal14}
\bibinfo{author}{\bibfnamefont{J.}~\bibnamefont{{Bramante}}},
  \bibinfo{author}{\bibfnamefont{K.}~\bibnamefont{{Fukushima}}},
  \bibinfo{author}{\bibfnamefont{J.}~\bibnamefont{{Kumar}}}, \bibnamefont{and}
  \bibinfo{author}{\bibfnamefont{E.}~\bibnamefont{{Stopnitzky}}},
  \bibinfo{journal}{\prd} \textbf{\bibinfo{volume}{89}}, \bibinfo{eid}{015010}
  (\bibinfo{year}{2014}), \eprint{1310.3509}.

\bibitem[{\citenamefont{{Albuquerque} et~al.}(2014)\citenamefont{{Albuquerque},
  {P{\'e}rez de los Heros}, and {Robertson}}}]{albuquerque_etal14}
\bibinfo{author}{\bibfnamefont{I.~F.~M.} \bibnamefont{{Albuquerque}}},
  \bibinfo{author}{\bibfnamefont{C.}~\bibnamefont{{P{\'e}rez de los Heros}}},
  \bibnamefont{and} \bibinfo{author}{\bibfnamefont{D.~S.}
  \bibnamefont{{Robertson}}}, \bibinfo{journal}{\jcap}
  \textbf{\bibinfo{volume}{2}}, \bibinfo{eid}{047} (\bibinfo{year}{2014}),
  \eprint{1312.0797}.

\bibitem[{\citenamefont{{Kaplinghat} et~al.}(2014)\citenamefont{{Kaplinghat},
  {Tulin}, and {Yu}}}]{kaplinghat_etal14}
\bibinfo{author}{\bibfnamefont{M.}~\bibnamefont{{Kaplinghat}}},
  \bibinfo{author}{\bibfnamefont{S.}~\bibnamefont{{Tulin}}}, \bibnamefont{and}
  \bibinfo{author}{\bibfnamefont{H.-B.} \bibnamefont{{Yu}}},
  \bibinfo{journal}{\prd} \textbf{\bibinfo{volume}{89}}, \bibinfo{eid}{035009}
  (\bibinfo{year}{2014}), \eprint{1310.7945}.

\bibitem[{\citenamefont{{Chen} et~al.}(2014)\citenamefont{{Chen}, {Lee}, {Lin},
  and {Lin}}}]{chen_etal14}
\bibinfo{author}{\bibfnamefont{C.-S.} \bibnamefont{{Chen}}},
  \bibinfo{author}{\bibfnamefont{F.-F.} \bibnamefont{{Lee}}},
  \bibinfo{author}{\bibfnamefont{G.-L.} \bibnamefont{{Lin}}}, \bibnamefont{and}
  \bibinfo{author}{\bibfnamefont{Y.-H.} \bibnamefont{{Lin}}},
  \bibinfo{journal}{\jcap} \textbf{\bibinfo{volume}{10}}, \bibinfo{eid}{049}
  (\bibinfo{year}{2014}), \eprint{1408.5471}.

\bibitem[{\citenamefont{{Feng} et~al.}(2016)\citenamefont{{Feng}, {Smolinsky},
  and {Tanedo}}}]{feng_etal16}
\bibinfo{author}{\bibfnamefont{J.~L.} \bibnamefont{{Feng}}},
  \bibinfo{author}{\bibfnamefont{J.}~\bibnamefont{{Smolinsky}}},
  \bibnamefont{and} \bibinfo{author}{\bibfnamefont{P.}~\bibnamefont{{Tanedo}}},
  \bibinfo{journal}{\prd} \textbf{\bibinfo{volume}{93}}, \bibinfo{eid}{115036}
  (\bibinfo{year}{2016}).

\bibitem[{\citenamefont{{Catena} and {Widmark}}(2016)}]{catena_widmark16}
\bibinfo{author}{\bibfnamefont{R.}~\bibnamefont{{Catena}}} \bibnamefont{and}
  \bibinfo{author}{\bibfnamefont{A.}~\bibnamefont{{Widmark}}},
  \bibinfo{journal}{\jcap} \textbf{\bibinfo{volume}{12}}, \bibinfo{eid}{016}
  (\bibinfo{year}{2016}), \eprint{1609.04825}.

\bibitem[{\citenamefont{Markevitch et~al.}(2004)\citenamefont{Markevitch,
  Gonzalez, Clowe, Vikhlinin, David, Forman, Jones, Murray, and
  Tucker}}]{Markevitch:2003at}
\bibinfo{author}{\bibfnamefont{M.}~\bibnamefont{Markevitch}},
  \bibinfo{author}{\bibfnamefont{A.~H.} \bibnamefont{Gonzalez}},
  \bibinfo{author}{\bibfnamefont{D.}~\bibnamefont{Clowe}},
  \bibinfo{author}{\bibfnamefont{A.}~\bibnamefont{Vikhlinin}},
  \bibinfo{author}{\bibfnamefont{L.}~\bibnamefont{David}},
  \bibinfo{author}{\bibfnamefont{W.}~\bibnamefont{Forman}},
  \bibinfo{author}{\bibfnamefont{C.}~\bibnamefont{Jones}},
  \bibinfo{author}{\bibfnamefont{S.}~\bibnamefont{Murray}}, \bibnamefont{and}
  \bibinfo{author}{\bibfnamefont{W.}~\bibnamefont{Tucker}},
  \bibinfo{journal}{Astrophys. J.} \textbf{\bibinfo{volume}{606}},
  \bibinfo{pages}{819} (\bibinfo{year}{2004}), \eprint{astro-ph/0309303}.

\bibitem[{\citenamefont{Zavala et~al.}(2013)\citenamefont{Zavala, Vogelsberger,
  and Walker}}]{Zavala:2012us}
\bibinfo{author}{\bibfnamefont{J.}~\bibnamefont{Zavala}},
  \bibinfo{author}{\bibfnamefont{M.}~\bibnamefont{Vogelsberger}},
  \bibnamefont{and} \bibinfo{author}{\bibfnamefont{M.~G.}
  \bibnamefont{Walker}}, \bibinfo{journal}{Monthly Notices of the Royal
  Astronomical Society: Letters} \textbf{\bibinfo{volume}{431}},
  \bibinfo{pages}{L20} (\bibinfo{year}{2013}), \eprint{1211.6426}.

\bibitem[{\citenamefont{Rocha et~al.}(2013)\citenamefont{Rocha, Peter, Bullock,
  Kaplinghat, Garrison-Kimmel, Onorbe, and Moustakas}}]{Rocha:2012jg}
\bibinfo{author}{\bibfnamefont{M.}~\bibnamefont{Rocha}},
  \bibinfo{author}{\bibfnamefont{A.~H.~G.} \bibnamefont{Peter}},
  \bibinfo{author}{\bibfnamefont{J.~S.} \bibnamefont{Bullock}},
  \bibinfo{author}{\bibfnamefont{M.}~\bibnamefont{Kaplinghat}},
  \bibinfo{author}{\bibfnamefont{S.}~\bibnamefont{Garrison-Kimmel}},
  \bibinfo{author}{\bibfnamefont{J.}~\bibnamefont{Onorbe}}, \bibnamefont{and}
  \bibinfo{author}{\bibfnamefont{L.~A.} \bibnamefont{Moustakas}},
  \bibinfo{journal}{Mon. Not. Roy. Astron. Soc.}
  \textbf{\bibinfo{volume}{430}}, \bibinfo{pages}{81} (\bibinfo{year}{2013}),
  \eprint{1208.3025}.

\bibitem[{\citenamefont{Peter et~al.}(2013)\citenamefont{Peter, Rocha, Bullock,
  and Kaplinghat}}]{Peter:2012jh}
\bibinfo{author}{\bibfnamefont{A.~H.~G.} \bibnamefont{Peter}},
  \bibinfo{author}{\bibfnamefont{M.}~\bibnamefont{Rocha}},
  \bibinfo{author}{\bibfnamefont{J.~S.} \bibnamefont{Bullock}},
  \bibnamefont{and}
  \bibinfo{author}{\bibfnamefont{M.}~\bibnamefont{Kaplinghat}},
  \bibinfo{journal}{Mon. Not. Roy. Astron. Soc.}
  \textbf{\bibinfo{volume}{430}}, \bibinfo{pages}{105} (\bibinfo{year}{2013}),
  \eprint{1208.3026}.

\bibitem[{\citenamefont{Kahlhoefer et~al.}(2014)\citenamefont{Kahlhoefer,
  Schmidt-Hoberg, Frandsen, and Sarkar}}]{Kahlhoefer:2013dca}
\bibinfo{author}{\bibfnamefont{F.}~\bibnamefont{Kahlhoefer}},
  \bibinfo{author}{\bibfnamefont{K.}~\bibnamefont{Schmidt-Hoberg}},
  \bibinfo{author}{\bibfnamefont{M.~T.} \bibnamefont{Frandsen}},
  \bibnamefont{and} \bibinfo{author}{\bibfnamefont{S.}~\bibnamefont{Sarkar}},
  \bibinfo{journal}{Mon. Not. Roy. Astron. Soc.}
  \textbf{\bibinfo{volume}{437}}, \bibinfo{pages}{2865} (\bibinfo{year}{2014}),
  \eprint{1308.3419}.

\bibitem[{\citenamefont{Elbert et~al.}(2015)\citenamefont{Elbert, Bullock,
  Garrison-Kimmel, Rocha, O–orbe, and Peter}}]{Elbert:2014bma}
\bibinfo{author}{\bibfnamefont{O.~D.} \bibnamefont{Elbert}},
  \bibinfo{author}{\bibfnamefont{J.~S.} \bibnamefont{Bullock}},
  \bibinfo{author}{\bibfnamefont{S.}~\bibnamefont{Garrison-Kimmel}},
  \bibinfo{author}{\bibfnamefont{M.}~\bibnamefont{Rocha}},
  \bibinfo{author}{\bibfnamefont{J.}~\bibnamefont{O–orbe}}, \bibnamefont{and}
  \bibinfo{author}{\bibfnamefont{A.~H.~G.} \bibnamefont{Peter}},
  \bibinfo{journal}{Mon. Not. Roy. Astron. Soc.}
  \textbf{\bibinfo{volume}{453}}, \bibinfo{pages}{29} (\bibinfo{year}{2015}),
  \eprint{1412.1477}.

\bibitem[{\citenamefont{Feng et~al.}(2016{\natexlab{a}})\citenamefont{Feng,
  Smolinsky, and Tanedo}}]{Feng:2015hja}
\bibinfo{author}{\bibfnamefont{J.~L.} \bibnamefont{Feng}},
  \bibinfo{author}{\bibfnamefont{J.}~\bibnamefont{Smolinsky}},
  \bibnamefont{and} \bibinfo{author}{\bibfnamefont{P.}~\bibnamefont{Tanedo}},
  \bibinfo{journal}{Phys. Rev.} \textbf{\bibinfo{volume}{D93}},
  \bibinfo{pages}{015014} (\bibinfo{year}{2016}{\natexlab{a}}),
  \eprint{1509.07525}.

\bibitem[{\citenamefont{Del~Nobile et~al.}(2015)\citenamefont{Del~Nobile,
  Kaplinghat, and Yu}}]{DelNobile:2015uua}
\bibinfo{author}{\bibfnamefont{E.}~\bibnamefont{Del~Nobile}},
  \bibinfo{author}{\bibfnamefont{M.}~\bibnamefont{Kaplinghat}},
  \bibnamefont{and} \bibinfo{author}{\bibfnamefont{H.-B.} \bibnamefont{Yu}},
  \bibinfo{journal}{JCAP} \textbf{\bibinfo{volume}{1510}}, \bibinfo{pages}{055}
  (\bibinfo{year}{2015}), \eprint{1507.04007}.

\bibitem[{\citenamefont{Feng et~al.}(2016{\natexlab{b}})\citenamefont{Feng,
  Smolinsky, and Tanedo}}]{Feng:2016ijc}
\bibinfo{author}{\bibfnamefont{J.~L.} \bibnamefont{Feng}},
  \bibinfo{author}{\bibfnamefont{J.}~\bibnamefont{Smolinsky}},
  \bibnamefont{and} \bibinfo{author}{\bibfnamefont{P.}~\bibnamefont{Tanedo}},
  \bibinfo{journal}{Phys. Rev.} \textbf{\bibinfo{volume}{D93}},
  \bibinfo{pages}{115036} (\bibinfo{year}{2016}{\natexlab{b}}),
  \eprint{1602.01465}.

\bibitem[{\citenamefont{Dooley et~al.}(2016)\citenamefont{Dooley, Peter,
  Vogelsberger, Zavala, and Frebel}}]{Dooley:2016ajo}
\bibinfo{author}{\bibfnamefont{G.~A.} \bibnamefont{Dooley}},
  \bibinfo{author}{\bibfnamefont{A.~H.~G.} \bibnamefont{Peter}},
  \bibinfo{author}{\bibfnamefont{M.}~\bibnamefont{Vogelsberger}},
  \bibinfo{author}{\bibfnamefont{J.}~\bibnamefont{Zavala}}, \bibnamefont{and}
  \bibinfo{author}{\bibfnamefont{A.}~\bibnamefont{Frebel}},
  \bibinfo{journal}{Mon. Not. Roy. Astron. Soc.}
  \textbf{\bibinfo{volume}{461}}, \bibinfo{pages}{710} (\bibinfo{year}{2016}),
  \eprint{1603.08919}.

\bibitem[{\citenamefont{Kim et~al.}(2016)\citenamefont{Kim, Peter, and
  Wittman}}]{Kim:2016ujt}
\bibinfo{author}{\bibfnamefont{S.~Y.} \bibnamefont{Kim}},
  \bibinfo{author}{\bibfnamefont{A.~H.~G.} \bibnamefont{Peter}},
  \bibnamefont{and} \bibinfo{author}{\bibfnamefont{D.}~\bibnamefont{Wittman}}
  (\bibinfo{year}{2016}), \eprint{1608.08630}.

\bibitem[{\citenamefont{Bringmann et~al.}(2017)\citenamefont{Bringmann,
  Kahlhoefer, Schmidt-Hoberg, and Walia}}]{Bringmann:2016din}
\bibinfo{author}{\bibfnamefont{T.}~\bibnamefont{Bringmann}},
  \bibinfo{author}{\bibfnamefont{F.}~\bibnamefont{Kahlhoefer}},
  \bibinfo{author}{\bibfnamefont{K.}~\bibnamefont{Schmidt-Hoberg}},
  \bibnamefont{and} \bibinfo{author}{\bibfnamefont{P.}~\bibnamefont{Walia}},
  \bibinfo{journal}{Phys. Rev. Lett.} \textbf{\bibinfo{volume}{118}},
  \bibinfo{pages}{141802} (\bibinfo{year}{2017}), \eprint{1612.00845}.

\bibitem[{\citenamefont{{Turner}}(1988)}]{turner88}
\bibinfo{author}{\bibfnamefont{M.~S.} \bibnamefont{{Turner}}},
  \bibinfo{journal}{Physical Review Letters} \textbf{\bibinfo{volume}{60}},
  \bibinfo{pages}{1797} (\bibinfo{year}{1988}).

\bibitem[{\citenamefont{{Raffelt}}(1996)}]{raffelt96_book}
\bibinfo{author}{\bibfnamefont{G.~G.} \bibnamefont{{Raffelt}}},
  \emph{\bibinfo{title}{{Stars as laboratories for fundamental physics : the
  astrophysics of neutrinos, axions, and other weakly interacting particles}}}
  (\bibinfo{publisher}{University of Chicago Press}, \bibinfo{year}{1996}).

\bibitem[{\citenamefont{{Dent} et~al.}(2012)\citenamefont{{Dent}, {Ferrer}, and
  {Krauss}}}]{dent_etal12}
\bibinfo{author}{\bibfnamefont{J.~B.} \bibnamefont{{Dent}}},
  \bibinfo{author}{\bibfnamefont{F.}~\bibnamefont{{Ferrer}}}, \bibnamefont{and}
  \bibinfo{author}{\bibfnamefont{L.~M.} \bibnamefont{{Krauss}}},
  \bibinfo{journal}{ArXiv e-prints}  (\bibinfo{year}{2012}),
  \eprint{1201.2683}.

\bibitem[{\citenamefont{{Rrapaj} and {Reddy}}(2016)}]{rrapaj_reddy16}
\bibinfo{author}{\bibfnamefont{E.}~\bibnamefont{{Rrapaj}}} \bibnamefont{and}
  \bibinfo{author}{\bibfnamefont{S.}~\bibnamefont{{Reddy}}},
  \bibinfo{journal}{\prc} \textbf{\bibinfo{volume}{94}}, \bibinfo{eid}{045805}
  (\bibinfo{year}{2016}), \eprint{1511.09136}.

\bibitem[{\citenamefont{{Hardy} and {Lasenby}}(2017)}]{hardy_lasenby17}
\bibinfo{author}{\bibfnamefont{E.}~\bibnamefont{{Hardy}}} \bibnamefont{and}
  \bibinfo{author}{\bibfnamefont{R.}~\bibnamefont{{Lasenby}}},
  \bibinfo{journal}{Journal of High Energy Physics}
  \textbf{\bibinfo{volume}{2}}, \bibinfo{eid}{33} (\bibinfo{year}{2017}),
  \eprint{1611.05852}.

\bibitem[{\citenamefont{Kazanas et~al.}(2014)\citenamefont{Kazanas, Mohapatra,
  Nussinov, Teplitz, and Zhang}}]{Kazanas:2014mca}
\bibinfo{author}{\bibfnamefont{D.}~\bibnamefont{Kazanas}},
  \bibinfo{author}{\bibfnamefont{R.~N.} \bibnamefont{Mohapatra}},
  \bibinfo{author}{\bibfnamefont{S.}~\bibnamefont{Nussinov}},
  \bibinfo{author}{\bibfnamefont{V.~L.} \bibnamefont{Teplitz}},
  \bibnamefont{and} \bibinfo{author}{\bibfnamefont{Y.}~\bibnamefont{Zhang}},
  \bibinfo{journal}{Nucl. Phys.} \textbf{\bibinfo{volume}{B890}},
  \bibinfo{pages}{17} (\bibinfo{year}{2014}), \eprint{1410.0221}.

\bibitem[{\citenamefont{Zhang}(2014)}]{Zhang:2014wra}
\bibinfo{author}{\bibfnamefont{Y.}~\bibnamefont{Zhang}},
  \bibinfo{journal}{JCAP} \textbf{\bibinfo{volume}{1411}}, \bibinfo{pages}{042}
  (\bibinfo{year}{2014}), \eprint{1404.7172}.

\bibitem[{\citenamefont{Chang et~al.}(2017)\citenamefont{Chang, Essig, and
  McDermott}}]{Chang:2016ntp}
\bibinfo{author}{\bibfnamefont{J.~H.} \bibnamefont{Chang}},
  \bibinfo{author}{\bibfnamefont{R.}~\bibnamefont{Essig}}, \bibnamefont{and}
  \bibinfo{author}{\bibfnamefont{S.~D.} \bibnamefont{McDermott}},
  \bibinfo{journal}{JHEP} \textbf{\bibinfo{volume}{01}}, \bibinfo{pages}{107}
  (\bibinfo{year}{2017}), \eprint{1611.03864}.

\bibitem[{\citenamefont{{Bjorken} et~al.}(2009)\citenamefont{{Bjorken},
  {Essig}, {Schuster}, and {Toro}}}]{bjorken_etal09}
\bibinfo{author}{\bibfnamefont{J.~D.} \bibnamefont{{Bjorken}}},
  \bibinfo{author}{\bibfnamefont{R.}~\bibnamefont{{Essig}}},
  \bibinfo{author}{\bibfnamefont{P.}~\bibnamefont{{Schuster}}},
  \bibnamefont{and} \bibinfo{author}{\bibfnamefont{N.}~\bibnamefont{{Toro}}},
  \bibinfo{journal}{\prd} \textbf{\bibinfo{volume}{80}}, \bibinfo{eid}{075018}
  (\bibinfo{year}{2009}), \eprint{0906.0580}.

\bibitem[{\citenamefont{{Kaplinghat} et~al.}(2013)\citenamefont{{Kaplinghat},
  {Tulin}, and {Yu}}}]{kaplinghat_etal13_whitepaper}
\bibinfo{author}{\bibfnamefont{M.}~\bibnamefont{{Kaplinghat}}},
  \bibinfo{author}{\bibfnamefont{S.}~\bibnamefont{{Tulin}}}, \bibnamefont{and}
  \bibinfo{author}{\bibfnamefont{H.-B.} \bibnamefont{{Yu}}},
  \bibinfo{journal}{ArXiv e-prints}  (\bibinfo{year}{2013}),
  \eprint{1308.0618}.

\bibitem[{\citenamefont{Hahn}(2005)}]{Hahn:2004fe}
\bibinfo{author}{\bibfnamefont{T.}~\bibnamefont{Hahn}},
  \bibinfo{journal}{Comput. Phys. Commun.} \textbf{\bibinfo{volume}{168}},
  \bibinfo{pages}{78} (\bibinfo{year}{2005}), \eprint{hep-ph/0404043}.

\bibitem[{\citenamefont{{Perego} et~al.}(2015)\citenamefont{{Perego}, {Hempel},
  {Fr{\"o}hlich}, {Ebinger}, {Eichler}, {Casanova}, {Liebend{\"o}rfer}, and
  {Thielemann}}}]{perego15}
\bibinfo{author}{\bibfnamefont{A.}~\bibnamefont{{Perego}}},
  \bibinfo{author}{\bibfnamefont{M.}~\bibnamefont{{Hempel}}},
  \bibinfo{author}{\bibfnamefont{C.}~\bibnamefont{{Fr{\"o}hlich}}},
  \bibinfo{author}{\bibfnamefont{K.}~\bibnamefont{{Ebinger}}},
  \bibinfo{author}{\bibfnamefont{M.}~\bibnamefont{{Eichler}}},
  \bibinfo{author}{\bibfnamefont{J.}~\bibnamefont{{Casanova}}},
  \bibinfo{author}{\bibfnamefont{M.}~\bibnamefont{{Liebend{\"o}rfer}}},
  \bibnamefont{and} \bibinfo{author}{\bibfnamefont{F.-K.}
  \bibnamefont{{Thielemann}}}, \bibinfo{journal}{\apj}
  \textbf{\bibinfo{volume}{806}}, \bibinfo{eid}{275} (\bibinfo{year}{2015}),
  \eprint{1501.02845}.

\bibitem[{\citenamefont{{Farmer} et~al.}(2016)\citenamefont{{Farmer}, {Fields},
  {Petermann}, {Dessart}, {Cantiello}, {Paxton}, and {Timmes}}}]{farmer16}
\bibinfo{author}{\bibfnamefont{R.}~\bibnamefont{{Farmer}}},
  \bibinfo{author}{\bibfnamefont{C.~E.} \bibnamefont{{Fields}}},
  \bibinfo{author}{\bibfnamefont{I.}~\bibnamefont{{Petermann}}},
  \bibinfo{author}{\bibfnamefont{L.}~\bibnamefont{{Dessart}}},
  \bibinfo{author}{\bibfnamefont{M.}~\bibnamefont{{Cantiello}}},
  \bibinfo{author}{\bibfnamefont{B.}~\bibnamefont{{Paxton}}}, \bibnamefont{and}
  \bibinfo{author}{\bibfnamefont{F.~X.} \bibnamefont{{Timmes}}},
  \bibinfo{journal}{\apjs} \textbf{\bibinfo{volume}{227}}, \bibinfo{eid}{22}
  (\bibinfo{year}{2016}), \eprint{1611.01207}.

\end{thebibliography}

\end{document}